\documentclass[
reprint,
superscriptaddress,
nofootinbib,
 amsmath,amssymb,
 aps,
prd,
floatfix,
]{revtex4-2}

\usepackage{graphicx}
\usepackage{dcolumn}
\usepackage{bm}
\usepackage{hyperref}

\usepackage[english]{babel}
\usepackage{xspace}
\usepackage[per-mode=symbol]{siunitx}
\usepackage[utf8]{inputenc}
\usepackage{multirow}
\usepackage{placeins}

\newcommand{\EE}{\ensuremath{E}\xspace}

\newcommand{\Nmu}{\ensuremath{N_\mu}\xspace}

\newcommand{\Nav}{\ensuremath{\langle N_\mu\rangle}\xspace}

\newcommand{\Emugtr}{\ensuremath{E_\mu > \SI{500}{\giga\eV}}\xspace}
\newcommand{\Sone}{\ensuremath{S_{125}}\xspace}
\newcommand{\tta}{\ensuremath{\theta}\xspace}
\newcommand{\lnA}{\ensuremath{\ln (A)}\xspace}

\newcommand{\sibyllpre}{Sibyll~2.1\xspace}

\newcommand{\epos}{EPOS-LHC\xspace}
\newcommand{\qgsjet}{QGSJet-II.04\xspace}
\newcommand{\corsika}{\textsc{Corsika}\xspace}
\newcommand{\proposal}{\textsc{Proposal}\xspace}
\newcommand{\fluka}{\textsc{Fluka}\xspace}
\newcommand{\geant}{\textsc{Geant4}\xspace}

\newcommand{\refeq}[1]{Eq.~(\ref{#1})}

\newcommand{\reffig}[1]{Fig.~\ref{#1}}
\newcommand{\reffigs}[2]{Figs.~\ref{#1}~and~\ref{#2}}

\newcommand{\refsec}[1]{Section~\ref{#1}}
\newcommand{\refsecs}[2]{Sections~\ref{#1}~and~\ref{#2}}

\newcommand{\refapp}[1]{Appendix~\ref{#1}}

\newcommand{\refref}[1]{Ref.~\cite{#1}}

\begin{document}

\title{Measurement of the mean number of muons with energies above 500 GeV in air showers detected with the IceCube Neutrino Observatory}

\affiliation{III. Physikalisches Institut, RWTH Aachen University, D-52056 Aachen, Germany}
\affiliation{Department of Physics, University of Adelaide, Adelaide, 5005, Australia}
\affiliation{Dept. of Physics and Astronomy, University of Alaska Anchorage, 3211 Providence Dr., Anchorage, AK 99508, USA}
\affiliation{School of Physics and Center for Relativistic Astrophysics, Georgia Institute of Technology, Atlanta, GA 30332, USA}
\affiliation{Dept. of Physics, Southern University, Baton Rouge, LA 70813, USA}
\affiliation{Dept. of Physics, University of California, Berkeley, CA 94720, USA}
\affiliation{Lawrence Berkeley National Laboratory, Berkeley, CA 94720, USA}
\affiliation{Institut f{\"u}r Physik, Humboldt-Universit{\"a}t zu Berlin, D-12489 Berlin, Germany}
\affiliation{Fakult{\"a}t f{\"u}r Physik {\&} Astronomie, Ruhr-Universit{\"a}t Bochum, D-44780 Bochum, Germany}
\affiliation{Universit{\'e} Libre de Bruxelles, Science Faculty CP230, B-1050 Brussels, Belgium}
\affiliation{Vrije Universiteit Brussel (VUB), Dienst ELEM, B-1050 Brussels, Belgium}
\affiliation{Dept. of Physics, Simon Fraser University, Burnaby, BC V5A 1S6, Canada}
\affiliation{Department of Physics and Laboratory for Particle Physics and Cosmology, Harvard University, Cambridge, MA 02138, USA}
\affiliation{Dept. of Physics, Massachusetts Institute of Technology, Cambridge, MA 02139, USA}
\affiliation{Dept. of Physics and The International Center for Hadron Astrophysics, Chiba University, Chiba 263-8522, Japan}
\affiliation{Department of Physics, Loyola University Chicago, Chicago, IL 60660, USA}
\affiliation{Dept. of Physics and Astronomy, University of Canterbury, Private Bag 4800, Christchurch, New Zealand}
\affiliation{Dept. of Physics, University of Maryland, College Park, MD 20742, USA}
\affiliation{Dept. of Astronomy, Ohio State University, Columbus, OH 43210, USA}
\affiliation{Dept. of Physics and Center for Cosmology and Astro-Particle Physics, Ohio State University, Columbus, OH 43210, USA}
\affiliation{Niels Bohr Institute, University of Copenhagen, DK-2100 Copenhagen, Denmark}
\affiliation{Dept. of Physics, TU Dortmund University, D-44221 Dortmund, Germany}
\affiliation{Dept. of Physics and Astronomy, Michigan State University, East Lansing, MI 48824, USA}
\affiliation{Dept. of Physics, University of Alberta, Edmonton, Alberta, T6G 2E1, Canada}
\affiliation{Erlangen Centre for Astroparticle Physics, Friedrich-Alexander-Universit{\"a}t Erlangen-N{\"u}rnberg, D-91058 Erlangen, Germany}
\affiliation{Physik-department, Technische Universit{\"a}t M{\"u}nchen, D-85748 Garching, Germany}
\affiliation{D{\'e}partement de physique nucl{\'e}aire et corpusculaire, Universit{\'e} de Gen{\`e}ve, CH-1211 Gen{\`e}ve, Switzerland}
\affiliation{Dept. of Physics and Astronomy, University of Gent, B-9000 Gent, Belgium}
\affiliation{Dept. of Physics and Astronomy, University of California, Irvine, CA 92697, USA}
\affiliation{Karlsruhe Institute of Technology, Institute for Astroparticle Physics, D-76021 Karlsruhe, Germany}
\affiliation{Karlsruhe Institute of Technology, Institute of Experimental Particle Physics, D-76021 Karlsruhe, Germany}
\affiliation{Dept. of Physics, Engineering Physics, and Astronomy, Queen's University, Kingston, ON K7L 3N6, Canada}
\affiliation{Department of Physics {\&} Astronomy, University of Nevada, Las Vegas, NV 89154, USA}
\affiliation{Nevada Center for Astrophysics, University of Nevada, Las Vegas, NV 89154, USA}
\affiliation{Dept. of Physics and Astronomy, University of Kansas, Lawrence, KS 66045, USA}
\affiliation{Centre for Cosmology, Particle Physics and Phenomenology - CP3, Universit{\'e} catholique de Louvain, Louvain-la-Neuve, Belgium}
\affiliation{Department of Physics, Mercer University, Macon, GA 31207-0001, USA}
\affiliation{Dept. of Astronomy, University of Wisconsin{\textemdash}Madison, Madison, WI 53706, USA}
\affiliation{Dept. of Physics and Wisconsin IceCube Particle Astrophysics Center, University of Wisconsin{\textemdash}Madison, Madison, WI 53706, USA}
\affiliation{Institute of Physics, University of Mainz, Staudinger Weg 7, D-55099 Mainz, Germany}
\affiliation{Department of Physics, Marquette University, Milwaukee, WI 53201, USA}
\affiliation{Institut f{\"u}r Kernphysik, Universit{\"a}t M{\"u}nster, D-48149 M{\"u}nster, Germany}
\affiliation{Bartol Research Institute and Dept. of Physics and Astronomy, University of Delaware, Newark, DE 19716, USA}
\affiliation{Dept. of Physics, Yale University, New Haven, CT 06520, USA}
\affiliation{Columbia Astrophysics and Nevis Laboratories, Columbia University, New York, NY 10027, USA}
\affiliation{Dept. of Physics, University of Oxford, Parks Road, Oxford OX1 3PU, United Kingdom}
\affiliation{Dipartimento di Fisica e Astronomia Galileo Galilei, Universit{\`a} Degli Studi di Padova, I-35122 Padova PD, Italy}
\affiliation{Dept. of Physics, Drexel University, 3141 Chestnut Street, Philadelphia, PA 19104, USA}
\affiliation{Physics Department, South Dakota School of Mines and Technology, Rapid City, SD 57701, USA}
\affiliation{Dept. of Physics, University of Wisconsin, River Falls, WI 54022, USA}
\affiliation{Dept. of Physics and Astronomy, University of Rochester, Rochester, NY 14627, USA}
\affiliation{Department of Physics and Astronomy, University of Utah, Salt Lake City, UT 84112, USA}
\affiliation{Dept. of Physics, Chung-Ang University, Seoul 06974, Republic of Korea}
\affiliation{Oskar Klein Centre and Dept. of Physics, Stockholm University, SE-10691 Stockholm, Sweden}
\affiliation{Dept. of Physics and Astronomy, Stony Brook University, Stony Brook, NY 11794-3800, USA}
\affiliation{Dept. of Physics, Sungkyunkwan University, Suwon 16419, Republic of Korea}
\affiliation{Institute of Physics, Academia Sinica, Taipei, 11529, Taiwan}
\affiliation{Dept. of Physics and Astronomy, University of Alabama, Tuscaloosa, AL 35487, USA}
\affiliation{Dept. of Astronomy and Astrophysics, Pennsylvania State University, University Park, PA 16802, USA}
\affiliation{Dept. of Physics, Pennsylvania State University, University Park, PA 16802, USA}
\affiliation{Dept. of Physics and Astronomy, Uppsala University, Box 516, SE-75120 Uppsala, Sweden}
\affiliation{Dept. of Physics, University of Wuppertal, D-42119 Wuppertal, Germany}
\affiliation{Deutsches Elektronen-Synchrotron DESY, Platanenallee 6, D-15738 Zeuthen, Germany}

\author{R. Abbasi}
\affiliation{Department of Physics, Loyola University Chicago, Chicago, IL 60660, USA}
\author{M. Ackermann}
\affiliation{Deutsches Elektronen-Synchrotron DESY, Platanenallee 6, D-15738 Zeuthen, Germany}
\author{J. Adams}
\affiliation{Dept. of Physics and Astronomy, University of Canterbury, Private Bag 4800, Christchurch, New Zealand}
\author{S. K. Agarwalla}
\thanks{also at Institute of Physics, Sachivalaya Marg, Sainik School Post, Bhubaneswar 751005, India}
\affiliation{Dept. of Physics and Wisconsin IceCube Particle Astrophysics Center, University of Wisconsin{\textemdash}Madison, Madison, WI 53706, USA}
\author{J. A. Aguilar}
\affiliation{Universit{\'e} Libre de Bruxelles, Science Faculty CP230, B-1050 Brussels, Belgium}
\author{M. Ahlers}
\affiliation{Niels Bohr Institute, University of Copenhagen, DK-2100 Copenhagen, Denmark}
\author{J.M. Alameddine}
\affiliation{Dept. of Physics, TU Dortmund University, D-44221 Dortmund, Germany}
\author{N. M. Amin}
\affiliation{Bartol Research Institute and Dept. of Physics and Astronomy, University of Delaware, Newark, DE 19716, USA}
\author{K. Andeen}
\affiliation{Department of Physics, Marquette University, Milwaukee, WI 53201, USA}
\author{C. Arg{\"u}elles}
\affiliation{Department of Physics and Laboratory for Particle Physics and Cosmology, Harvard University, Cambridge, MA 02138, USA}
% ...
\date{\today}

\collaboration{IceCube Collaboration}
\noaffiliation


\begin{abstract}
    We present a measurement of the mean number of muons with energies larger than 500 GeV in near-vertical extensive air showers initiated by cosmic rays with primary energies between \SI{2.5}{\peta\eV} and \SI{100}{\peta\eV}. The measurement is based on events detected in coincidence between the surface and in-ice detectors of the IceCube Neutrino Observatory. Air showers are recorded on the surface by IceTop, while a bundle of high-energy muons (TeV muon) from the shower can subsequently produce a tracklike event in the IceCube in-ice array. Results are obtained assuming the hadronic interaction models Sibyll 2.1, QGSJet-II.04, and EPOS-LHC. The measured number of TeV muons is found to be in agreement with predictions from air-shower simulations. The results have also been compared to a measurement of low-energy muons by IceTop, indicating an inconsistency between the predictions for low- and high-energy muons in simulations based on the EPOS-LHC model.
\end{abstract}

\maketitle

\section{Introduction}\label{sec:intro}

Cosmic rays are ionized nuclei coming from outer space with energies extending up to \SI{e20}{\eV} and beyond.
A detailed knowledge of the properties of the cosmic-ray flux is important for understanding their origin. In addition to directly carrying information about the sources and propagation of cosmic rays, it is a major uncertainty in the calculation of the atmospheric neutrino flux, the main background for neutrino astronomy~\cite{Evans:2016obt, Yanez:2023lsy}.
The high energies of cosmic rays furthermore provide the opportunity to explore particle physics beyond the reach of human-made accelerators~\cite{Engel:2011zzb}.

While the energy spectrum of cosmic rays has been measured with high precision over many orders of magnitude, the mass composition remains uncertain. This results from the fact that cosmic rays with energies above several 100 TeV are observed indirectly through the extensive air showers they produce when entering the Earth's atmosphere. The properties of the primary cosmic rays can be inferred from several observables accessible through ground-based experiments. One such observable is the number of muons in the air shower, which can be probed with arrays of particle detectors. Muons originate predominantly from the decay of charged pions and kaons produced as part of the hadronic cascade, and, together with an independent estimate of the primary energy, their number can be used to infer the mass of the cosmic-ray nucleus~\cite{Kampert:2012mx}. The interpretation of the air-shower observations in terms of properties of the primary particle relies, however, on the accurate simulation of the development of the air shower in the atmosphere. These Monte Carlo simulations make use of phenomenological models describing the high-energy hadronic interactions in the shower, which cannot be calculated in the context of perturbative quantum chromodynamics. The models are tuned to accelerator data, but have to extrapolate to a phase space not accessible to these experiments, leading to significant uncertainties~\cite{Engel:2011zzb, Pierog:2019srb}.

Several observatories have performed measurements of the muon content of air showers.
A 2019 meta-analysis including data from eight observatories found a discrepancy between data and expectations from simulations with a significant dependence on the shower energy~\cite{EAS-MSU:2019kmv}, with the Pierre Auger Observatory reporting evidence for a deficit of muons in simulations based on post-LHC hadronic interaction models at the highest energies~\cite{PierreAuger:2014ucz, PierreAuger:2021qsd}.
This observation is commonly referred to as \textit{the Muon Puzzle}, its most plausible origin being the description of hadronic interactions in the air showers~\cite{Albrecht:2021cxw}.
The current experimental picture, however, is ambiguous; while some observatories report a discrepancy between measurements and simulations based on various hadronic interaction models, others do not observe any such disagreements~\cite{ArteagaVelazquez:2023fda}.
The observatories all operate in different regions of the parameter space, such as observation altitude and muon energy threshold, and a consistent picture has yet to emerge.
As a result, interpreting the muon measurements of different observatories in terms of the cosmic-ray mass composition leads to inconsistent results.
Resolving this situation, i.e. the Muon Puzzle and the possibly related inconsistencies between observatories, is considered one of the most pressing problems in high-energy cosmic-ray physics and motivates performing a variety of muon measurements to probe air-shower development and hadronic interactions under different conditions~\cite{Coleman:2022abf}.
Complementary information toward understanding the issues in the description of high-energy hadronic interactions will also be obtained by collider experiments, with those measuring in the forward region being particularly important for air-shower physics~\cite{Albrecht:2021cxw, Soldin:2023gox}.

The IceCube Neutrino Observatory~\cite{IceCube:2016zyt} is a particle detector located at the geographic South Pole in Antarctica. Its combination of a surface detector array, IceTop, and the large-volume in-ice detector that is located between \SI{1.5}{\km} and \SI{2.5}{\km} below, allows it to perform unique air-shower measurements. A measurement of the density of mainly low-energy muons detected at the surface with IceTop, often referred to as GeV muons, has been previously published~\cite{IceCubeCollaboration:2022tla}. The thick ice sheet covering the IceCube in-ice detector absorbs all muons with energies below several hundred GeV, allowing for a measurement of purely the high-energy muon content in air showers. These muons, which we refer to as TeV muons, are the subject of this article. The TeV muons are of special interest as they are predominantly produced in the early stages of the shower development. Their number depends on the energy and mass of the primary. The mass dependence is stronger than that of low-energy muons observed with surface arrays, making them also a particularly interesting observable for mass composition studies~\cite{Flaggs:2023exc}(see also \refapp{app:HE_mu}). In this work, we present a measurement of the mean number of muons with energies above \SI{500}{\giga\eV} in near-vertical air showers observed with both IceTop and the IceCube in-ice array in the primary energy range of \SI{2.5}{\peta\eV} to \SI{100}{\peta\eV}.

The article is structured as follows. In \refsec{sec:detectors}, we introduce the IceCube Neutrino Observatory and its operation as an extensive-air-shower detector. Following this, the experimental data and simulations used in the analysis are described in \refsec{sec:datasets}. \refsec{sec:analysis} describes in detail the analysis method, including neural network reconstructions of the primary cosmic-ray energy and TeV muon multiplicity, as well as the application of correction factors derived from simulations. The final results are presented in \refsec{sec:results}. Two appendices are included: \refapp{app:HE_mu} discusses the predictions for high-energy muons from simulations in detail; \refapp{app:checks} includes various demonstrations of the robustness of the analysis results.

\section{IceTop and IceCube In-Ice}\label{sec:detectors}

\begin{figure}[bt]
    \centering
    \includegraphics[width=0.7\linewidth]{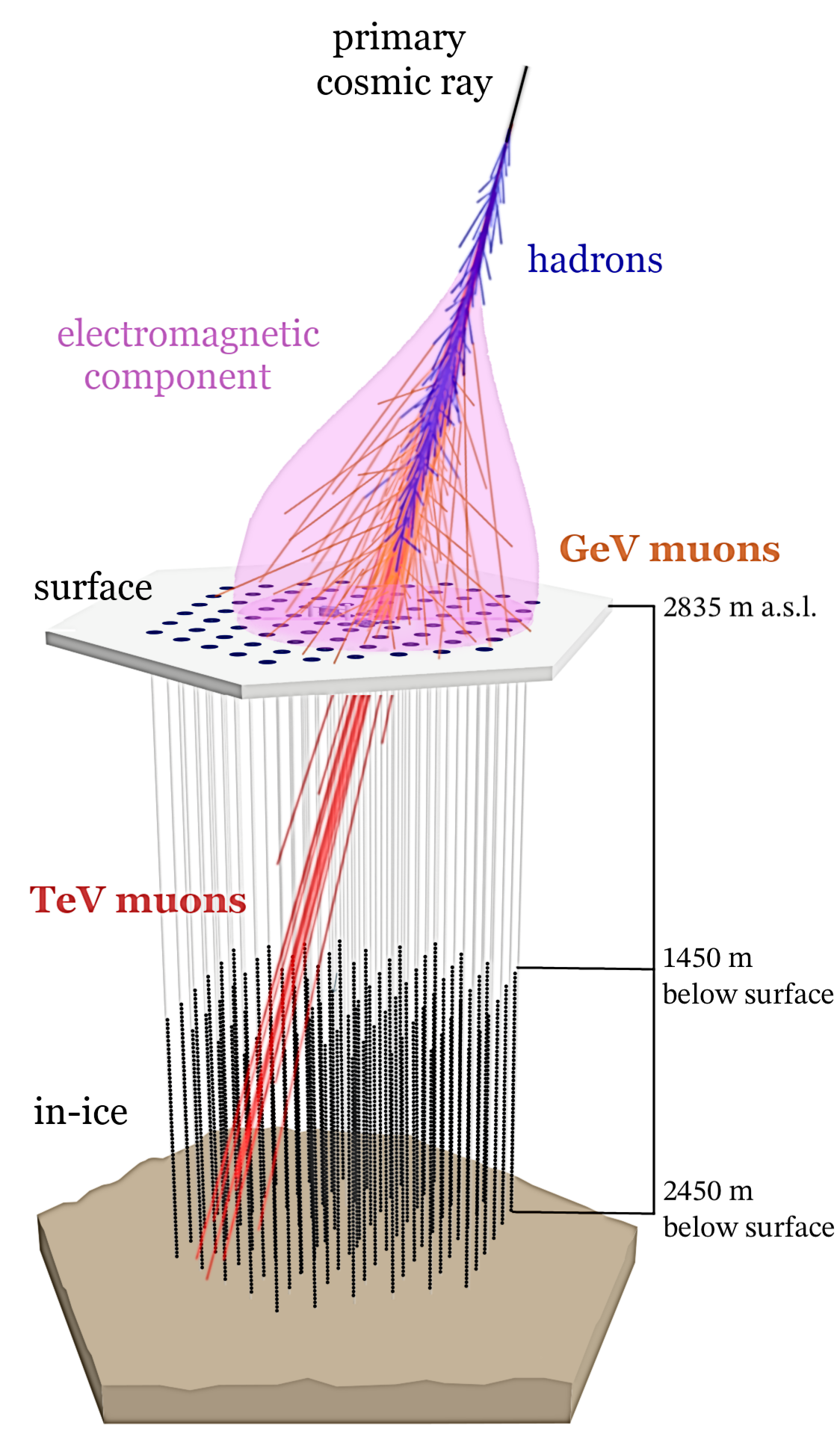}
    \vspace{-0.1cm}
    \caption{Schematic drawing of a cosmic-ray air shower observed in coincidence between IceTop, the surface component of the IceCube Neutrino Observatory, and the IceCube in-ice array.}
    \label{fig:icecube}
\end{figure}

The IceCube Neutrino Observatory, illustrated in \reffig{fig:icecube}, is a kilometer-scale detector located at the geographic South Pole, constructed mainly for the detection of high-energy neutrinos and cosmic rays~\cite{IceCube:2016zyt}. The IceCube in-ice array consists of 5160 digital optical modules (DOMs) deployed on 86 vertical strings at a depth between \SI{1450}{\m} and \SI{2450}{\m} with a horizontal spacing of \SI{125}{\m}, instrumenting a volume of one cubic kilometer of ice. Each DOM contains a photomultiplier tube sensitive to the Cherenkov photons produced by relativistic charged particles propagating through the ice. The surface component of the observatory, IceTop~\cite{IceCube:2012nn}, is an array consisting of 81 stations of two ice-Cherenkov tanks approximately following the in-ice string locations, covering an area of about one square kilometer. Each tank contains two DOMs to record the Cherenkov light produced by air-shower particles penetrating the ice in the tank. The lids of the tanks were deployed flush with the snow surface at the site. Since then, however, several meters of snow have accumulated on top of the array.

The IceTop array is situated at an altitude of \SI{2835}{\m} above sea level, corresponding to an atmospheric depth of about \SI{690}{\g\per\cm\squared}. IceTop signals are calibrated in terms of the typical signal produced by vertical muons penetrating the tanks, known as the vertical equivalent muon or VEM. The signals recorded in the tanks for an air shower whose impact point is located inside the array, are dominated by the electromagnetic shower component, except at large lateral distances, where a muonic signal component can be discerned. This feature has been used previously to measure the density of mainly low-energy ($\mathcal{O}$(GeV)) muons at the surface~\cite{IceCubeCollaboration:2022tla}. For air showers whose shower axis also passes through the volume of the in-ice detector, a narrow bundle of high-energy muons can be observed in coincidence with the surface signals. Muons need an energy of at least several hundred GeV to travel from the surface to the top of the in-ice array.
When requiring that the shower core is contained within the IceTop footprint, events with a coincident surface trigger and in-ice muon bundle are limited to zenith angles of $\theta \lesssim 38^\circ$.
This class of events has been used before to perform a measurement of the cosmic-ray mass composition~\cite{IceCube:2019hmk}.

The IceTop signals are typically used to reconstruct various parameters, such as the shower core position and direction, with a maximum likelihood method which fits the time and charge distributions~\cite{IceCube:2012nn}. This procedure also reconstructs the shower size $S_{125}$, the signal at a distance of \SI{125}{\m} from the shower axis, which is a proxy for the primary cosmic-ray energy. A simple exponential attenuation function is used during the reconstruction to take the impact of the snow coverage on top of a tank on the observed signal into account~\cite{IceCube:2013ftu}. Additional information about the primary particle and the air shower can be obtained from the in-ice detector, typically by studying the charge deposit of the high-energy muon bundle as it propagates through the ice.
One method, which has been used in previous analyses, consists of determining the energy loss in equidistant segments along the line of propagation of the bundle. As the photons detected by a DOM can originate from different positions along the track, an unfolding is required to obtain the contributions from energy losses in all track segments to the observed light distribution~\cite{IceCube:2013dkx}.
These established reconstruction methods are applied to all air-shower events included in the analysis presented in this paper.

\section{Datasets}\label{sec:datasets}

\subsection{Experimental dataset}

The analysis uses data collected between May 15, 2012 and May 2, 2013 with an effective live time of about 323 days. A total of \num{1216154} events with a reconstructed energy between \SI{2.5}{\peta\eV} and \SI{100}{\peta\eV} pass the selection criteria described below.

The event selection is aimed at air showers which trigger IceTop and have a coincident bundle of high-energy muons in the in-ice detector. To obtain an event sample with a high quality of air-shower reconstructions, cuts established in previous IceCube cosmic-ray analyses are applied~\cite{IceCube:2013ftu, IceCube:2019hmk}. Events must trigger at least five IceTop stations and at least one of the stations must have a signal greater than 6 VEM. The shower reconstruction is required to succeed, and the reconstructed shower core is required to be within a geometric boundary slightly smaller than the array. Furthermore, the station with the largest signal must not be on the edge of the array. For the in-ice part of the events, quality cuts are applied to the unfolded muon-bundle energy loss, requiring a successful reconstruction which has at least three reconstructed energy losses inside the detector volume that are nonzero (see also \refsec{sec:NN}). While this selection can include events up to about 38$^\circ$ in zenith, the analysis is further restricted to events with $\cos \theta > 0.95$ ($\theta \lesssim 18^\circ$). This simplifies the possible muon bundle geometries included in the analysis and ensures that all muon bundles have propagated through a similar amount of matter before reaching the detector, while still resulting in a systematics-dominated measurement.

 As will be discussed in \refsec{sec:NN}, the resolution of the cosmic-ray energy reconstruction method that is used in this analysis, defined as the standard deviation of $\log_{10} (E_\mathrm{reco}/E_\mathrm{true})$, is better than 0.1 above \SI{1}{\peta\eV} and about 0.05 above \SI{10}{\peta\eV}. The analysis only includes events above the threshold energy where the event selection reaches nearly full efficiency for all masses of cosmic-ray nuclei, i.e.~\SI{2.5}{\peta\eV} for the level of snow coverage on the IceTop array in the included time interval~\cite{IceCube:2019hmk}, and includes events of energies up to \SI{100}{\peta\eV}. Around the threshold energy, a core resolution better than \SI{12}{\m} and an angular resolution better than $0.8^\circ$ are obtained after quality cuts, improving to about \SI{5}{\m} and $0.4^\circ$ at \SI{30}{\peta\eV}~\cite{Verpoest:2022zya}.

\subsection{Simulated datasets}

Air shower simulations are produced with \corsika v7.3700~\cite{Heck:1998vt}, using an atmospheric model describing a typical South Pole atmosphere for the month of April.\footnote{This atmospheric model was included as a standard model in \corsika v7.4700 with identifier 33.} This model closely approximates the yearly average South Pole atmosphere~\cite{DeRidder:2019ofg}. Primary energies are sampled according to an $E^{-1}$ differential energy spectrum. Hadrons and (anti)muons are simulated down to energies of \SI{0.05}{\giga\eV}, while electrons/positrons and photons are included down to \SI{0.01}{\giga\eV} and \SI{0.002}{\giga\eV} respectively. 

The model used for hadronic interactions below a laboratory energy of \SI{80}{\giga\eV} is \fluka 2011.2c~\cite{Ferrari:2005zk, Bohlen:2014buj}. Datasets with different high-energy hadronic interaction models are included in the analysis. The main dataset is based on \sibyllpre~\cite{Ahn:2009wx} and includes four types of primary cosmic ray: p, He, O, and Fe. Two other datasets are based on \qgsjet~\cite{Ostapchenko:2010vb} and \epos~\cite{Pierog:2013ria} and include only proton and iron showers. \sibyllpre is a \textit{pre-LHC} model, while \qgsjet and \epos are \textit{post-LHC} models, taking into account high-energy data from the LHC.

The detector response to the shower particles is simulated using IceCube software including the entire hardware and data-acquisition chain, as in previous analyses~\cite{IceCube:2013ftu, IceCube:2019hmk}. Each \corsika shower is resampled 100 times uniformly over an energy-dependent area larger than the IceTop detector area. The observation level in \corsika is set to \SI{2837}{\m}, which is several meters above the top of the IceTop tanks to allow for the inclusion of snow accumulated on top of the tanks in the simulation. The shower particles obtained from \corsika are propagated from the observation level through layers of air and snow before their interactions in the tanks are simulated, all using the \geant~\cite{GEANT4:2002zbu} package. The simulated snow levels are those measured \textit{in situ} in October 2012. Only muons with energies higher than \SI{273}{\giga\eV} are considered for simulation of the in-ice detector response, including their interactions with the ice and the resulting detection of Cherenkov photons. The same reconstructions and event selection applied to the experimental data are applied to the simulated events.

Additional \corsika-only datasets are produced to obtain predictions of the multiplicity of muons with \Emugtr with reduced statistical uncertainty, for comparison to the final analysis results. Predictions are derived for p and Fe primaries using the models \sibyllpre, \qgsjet, and \epos. They are described in more detail in \refapp{app:HE_mu}.

\begin{figure*}
    \centering
    \raisebox{0mm}{\includegraphics[width=0.45\textwidth]{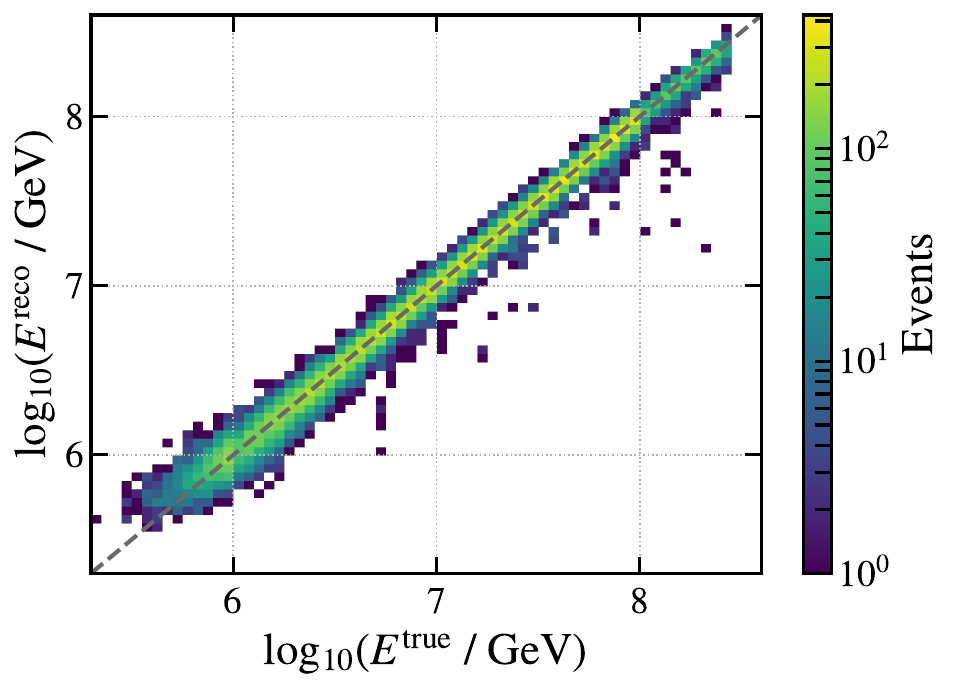}}\raisebox{-1mm}{\includegraphics[width=0.47\textwidth]{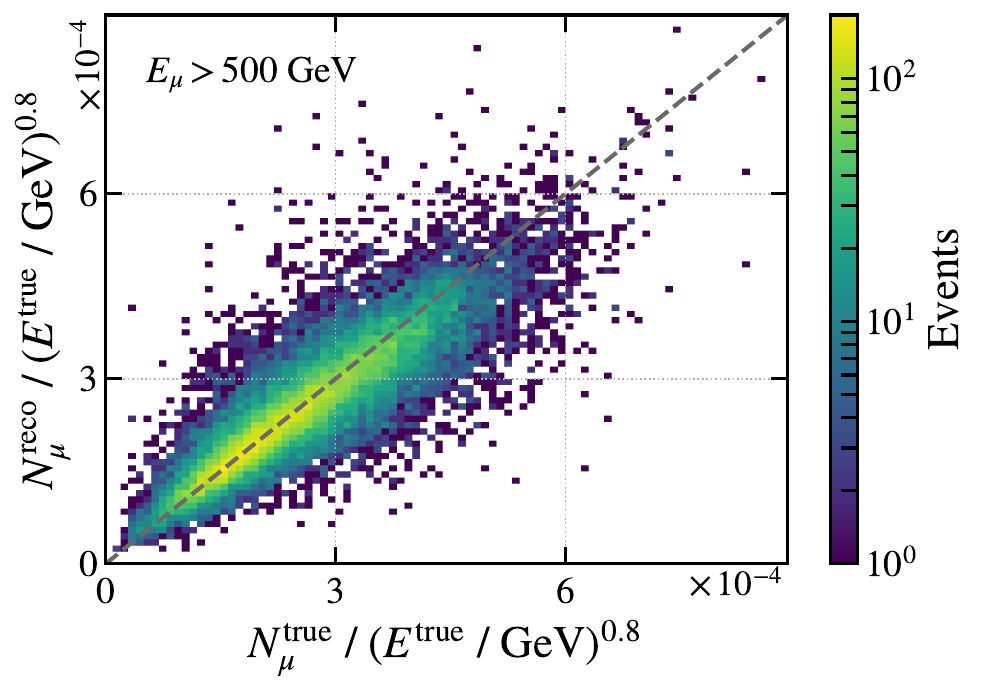}}
    \vspace{-0.1cm}
    \caption{Correlation plots showing the relation between the true and neural-network reconstructed values for primary cosmic-ray energy (left) and high-energy muon multiplicity (right). The muon multiplicity has been divided by $E^\beta$ to reduce the effect of its underlying energy dependence. The histograms include \sibyllpre simulations of all four mass groups (p, He, O, and Fe).}
    \label{fig:NN_correlation}
\end{figure*}

\section{Analysis}\label{sec:analysis}

The analysis presented in this paper determines the mean number of muons \Nmu with an energy \Emugtr in air showers with $\cos\theta > 0.95$ for primary cosmic-ray energies \EE between \SI{2.5}{\peta\eV} and \SI{100}{\peta\eV}, where \Nmu is defined at the surface.\footnote{Simulations of vertically down-going muons performed with \proposal~\cite{Koehne:2013gpa} show that muons with an energy of \SI{500}{\giga\eV} have a chance of about 75\% to propagate down to at least \SI{1450}{\m} in the ice. This increases to 95\% for muons with an energy of \SI{1}{\tera\eV}.} The analysis starts from the initial air-shower and muon-bundle energy-loss reconstructions and the event selection discussed in \refsecs{sec:detectors}{sec:datasets}. The reconstructed observables from IceTop and in-ice are used as inputs for a neural network reconstruction of the primary cosmic-ray energy and the high-energy muon number. Correction factors are subsequently derived from simulations and used to correct for mass-dependent biases in the determination of the mean muon number \Nav in bins of primary energy.

\subsection{Neural network reconstruction}\label{sec:NN}

A neural network model is used to relate the signals generated by the bundle of high-energy muons in the in-ice IceCube detector to the number of muons in the air shower with an energy greater than \SI{500}{\giga\eV} at the surface. In addition, the shower size \Sone reconstructed with IceTop provides the main sensitivity to the primary cosmic-ray energy. Various ways of combining the inputs in different neural networks were explored. For example, one could use only surface information for the reconstruction of the energy \EE and only in-ice information for the reconstruction of the muon number \Nmu, or one could utilize them in a single model for combined reconstruction. Some of these approaches are discussed in more detail in \refapp{app:checks}, demonstrating that the final results are invariant to such choices. We present in this section the approach that was used to obtain the nominal analysis results which are presented in \refsec{sec:results}.

The muon bundle energy loss is reconstructed in segments of \SI{20}{\m} along the shower axis, which is obtained from the IceTop reconstruction and extended to the in-ice detector. Any segment not contained in the detector volume is removed. Events which end up with less than three segments that have a nonzero reconstructed energy loss in the detector are discarded. The first segment in the detector will correspond to a different slant depth traveled by the bundle for events with different zenith angles.
To include this information, we create a fixed-length vector of reconstructed energy losses in such a way that each entry in the vector corresponds approximately to the same in-ice slant depth for all events, padding with zeros at the start or end of the vector based on the zenith angle. More specifically, a vertical event will have zeros at the end of the vector, while a more inclined event will have zeros at the start, in such a way that the first entry corresponds to a traveled distance of about \SI{1450}{\m} in the ice.
This vector of energy losses is used as input to a recurrent neural network, more specifically a bidirectional gated recurrent unit layer~\cite{chung2014empirical}, a common choice for sequential data. The output from the recurrent layer is fed into a fully connected (dense) layer combined with the shower size $S_{125}$ and zenith angle $\theta$ reconstructed using IceTop. The neural network subsequently outputs predictions for both $\log_{10}(\Nmu)$ and $\log_{10}(\EE)$.
The network is trained using a mean-squared error (MSE) loss function for each of the training targets, minimizing the combined loss function consisting of the sum of the two MSE losses.
The implementation of the neural network is based on the \textsc{Keras}~\cite{chollet2015keras} and \textsc{TensorFlow}~\cite{abadi2016tensorflow} software libraries.

\begin{figure*}
    \centering
    \includegraphics[width=0.42\textwidth]{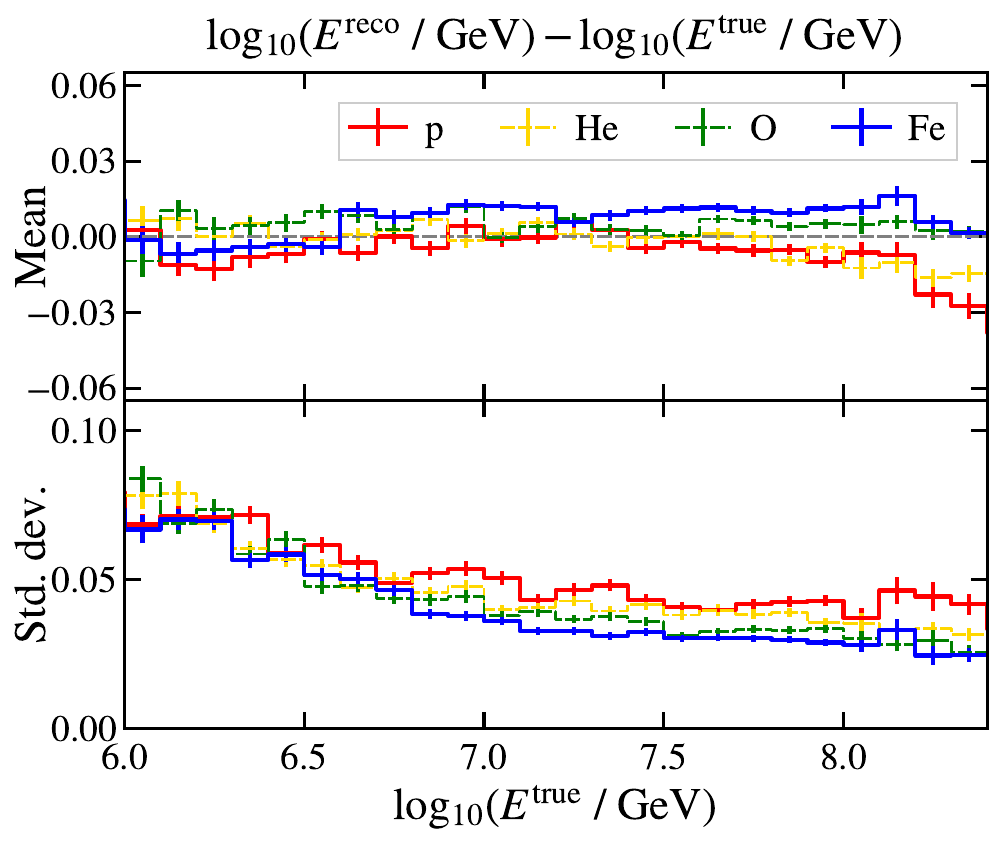}\qquad\includegraphics[width=0.42\textwidth]{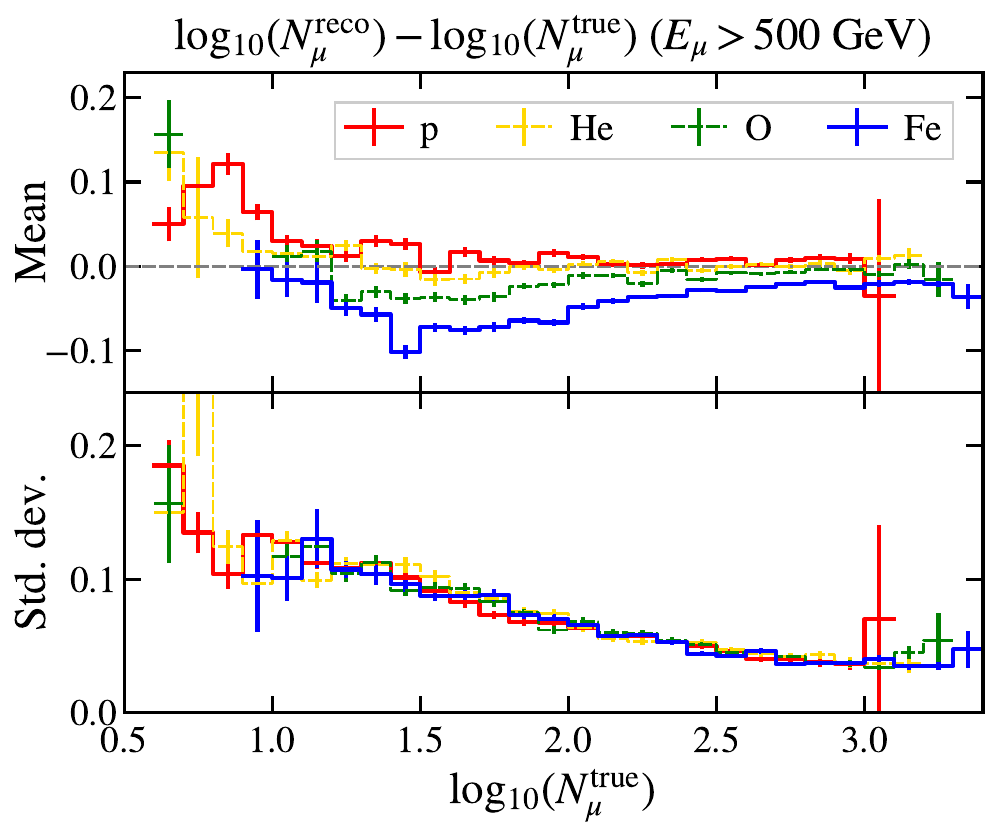}
    \vspace{-0.1cm}
    \caption{Bias and resolution of the neural network reconstructions, defined as the mean and standard deviation of the difference between the logarithms of the reconstructed and true values, shown separately for different primary cosmic-ray masses. Primary energy reconstruction is shown on the left, muon multiplicity (\Emugtr) reconstruction on the right. The analysis includes events between 6.4 and 8.0 in $\log_{10} (E/\mathrm{GeV})$, corresponding to $0.7 \lesssim \log_{10} (N_\mu) \lesssim 2.8$ for proton showers and $1.5 \lesssim \log_{10} (N_\mu) \lesssim 3.1$ for iron showers.}
    \label{fig:NN_reso}
\end{figure*}

The neural network is trained on \sibyllpre simulations with roughly equal amounts of the four primaries (p, He, O, and Fe), with true energies in the range $5.4 \leq \log_{10} (\EE\ /\ \mathrm{GeV}) \leq 8.4$, corresponding to a range for \Nmu of one muon up to about 2500. The quality of the reconstructions is examined by applying them to an independent test set. \reffig{fig:NN_correlation} shows the relations between the reconstructed and true values for \EE and \Nmu, combined for all primaries. As the high-energy muon multiplicity approximately grows as $E^\beta$ with $\beta \approx 0.8$, as discussed in \refapp{app:HE_mu}, the ratio $N_\mu/E^{0.8}$ has been plotted to prevent the correlation from simply reflecting the underlying energy dependence rather than the performance of the neural network. The bias and resolution of the reconstructions, defined as respectively the mean and the standard deviation of the logarithmic difference between reconstructed and true values, is shown in \reffig{fig:NN_reso} for the different primary types separately. For most of the energy range included in the analysis, i.e.~6.4 to 8.0 in $\log_{10} (E/\mathrm{GeV})$, the energy reconstruction has a resolution smaller than half the bin width of 0.1 in $\log_{10}(E)$, with some mass dependence in both the bias and the resolution. The muon number reconstruction, on the other hand, has a nearly mass-independent resolution, but has a clear mass-dependent bias.\footnote{Training two separate neural networks, one for \EE reconstruction using only IceTop information, and one for \Nmu reconstruction using only in-ice information, changes this behavior. In that case, a larger mass dependence in the \EE reconstruction bias is obtained, while the \Nmu network produces a reduced mass-dependent bias. The combined neural network presented in the main body of this work allows, however, for the more precise correction method for the resulting biases discussed in \refsec{sec:correction}. See \refapp{app:checks} for a more detailed discussion.} The number of muons with \Emugtr in simulations inside the primary energy range used in the analysis is between 5 and about 670 for proton showers and between 30 and about 1330 for iron showers.

\subsection{Monte Carlo correction}\label{sec:correction}

\begin{figure}
    \centering
    \includegraphics[width=0.9\linewidth, trim = 0 0.8em 0 0, clip]{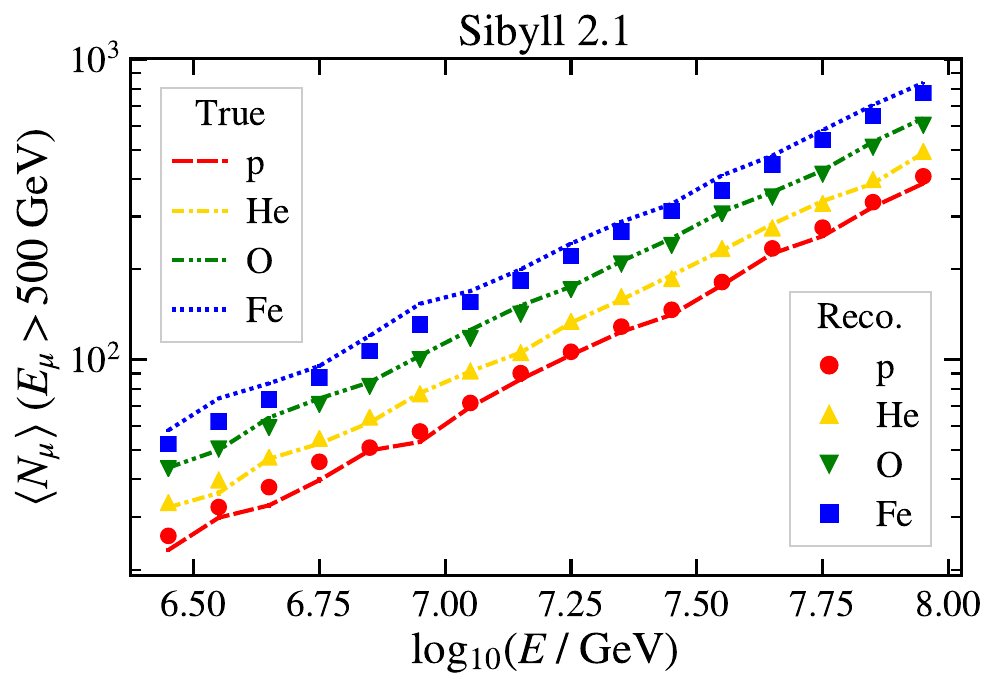}

    \includegraphics[width=0.9\linewidth, trim = 0 0.8em 0 0, clip]{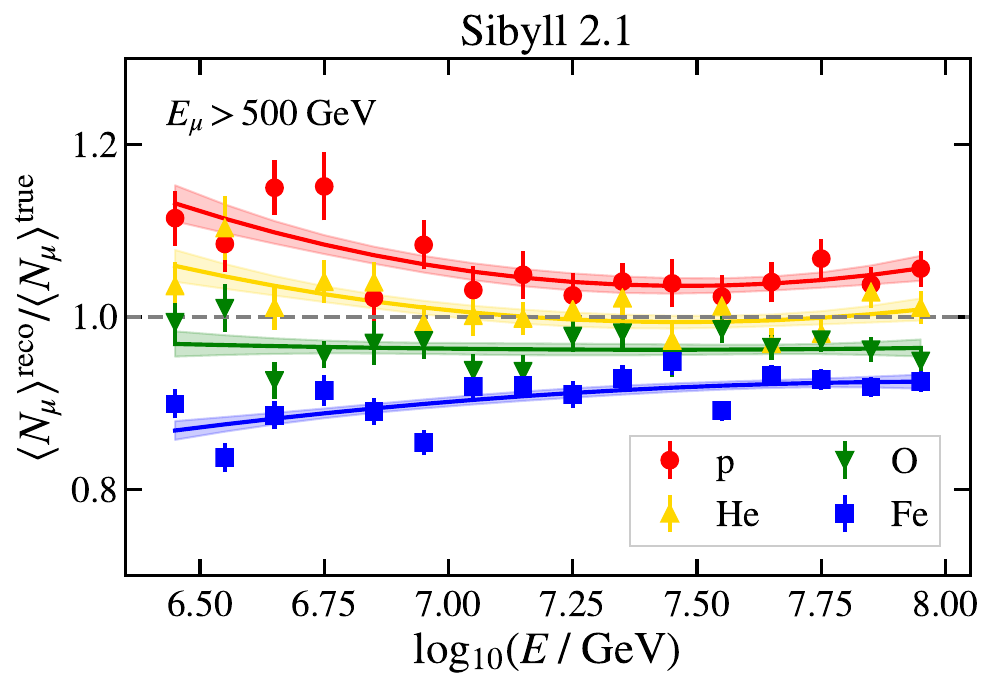}
    \vspace{-0.1cm}
    \caption{Top: comparison between the average reconstructed TeV muon number in bins of reconstructed primary energy and the true muon number in bins of true energy in \sibyllpre simulation for four different primaries. Bottom: ratio of the reconstructed and true values from the top plot, fitted with quadratic functions defining correction factors based on \sibyllpre.}
    \label{fig:reco_v_true}
\end{figure}

The analysis determines the average high-energy muon number \Nav in bins of primary cosmic-ray energy $E$ based on the event-by-event reconstructions described above. The accuracy with which it can be obtained is tested based on simulations. In \reffig{fig:reco_v_true}, the average of the reconstructed muon number values is shown in bins of reconstructed energy for different primaries, compared to the true average muon number in bins of true simulated energy. These values were determined from the subset of the \sibyllpre dataset which was not used in the training of the neural network. While the energy dependence is captured rather well for all masses, systematic offsets can be seen between the reconstructed and true values. The bottom panel of the figure shows the ratio between the reconstructed and true values, indicating biases up to about $\pm 15\%$ with a clear mass dependence. The systematic overestimation for proton and underestimation for iron arises from the tendency of neural networks trained with MSE loss to predict values closer to the average behavior of the training data. The ratios are fit with quadratic functions for each mass separately, defining the correction factors that will be applied to the results. We note that the offsets shown in these plots result from both the imperfections of the muon multiplicity reconstruction, which has the largest impact, as well as from the energy reconstruction as a result of bin migration, which is a smaller effect.
 
\begin{figure*}
    \centering
    \includegraphics[width=0.45\textwidth]{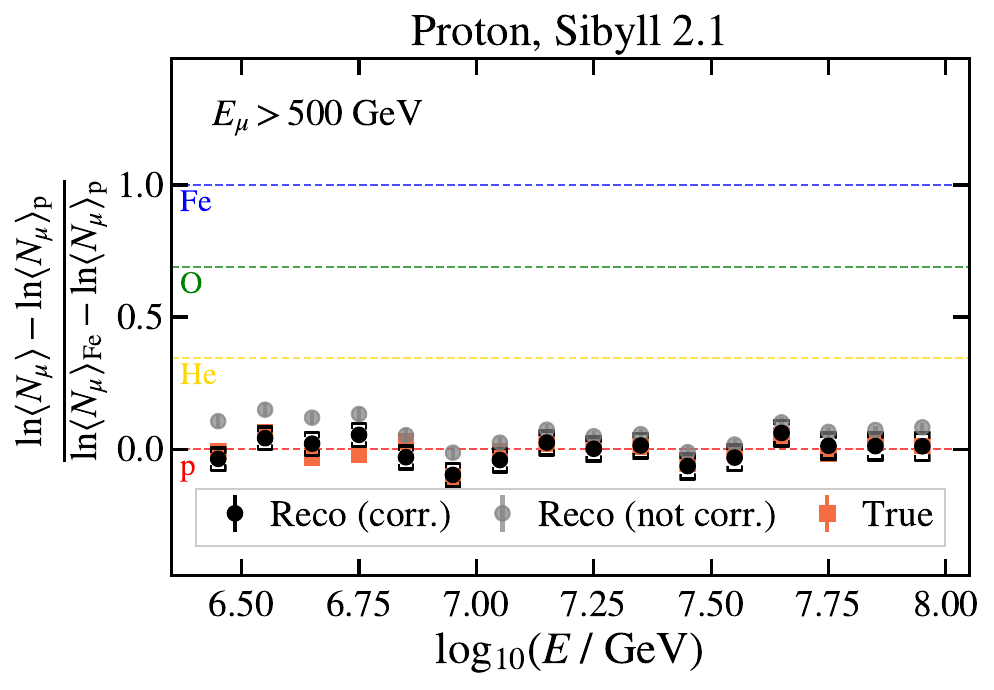}\qquad\includegraphics[width=0.45\textwidth]{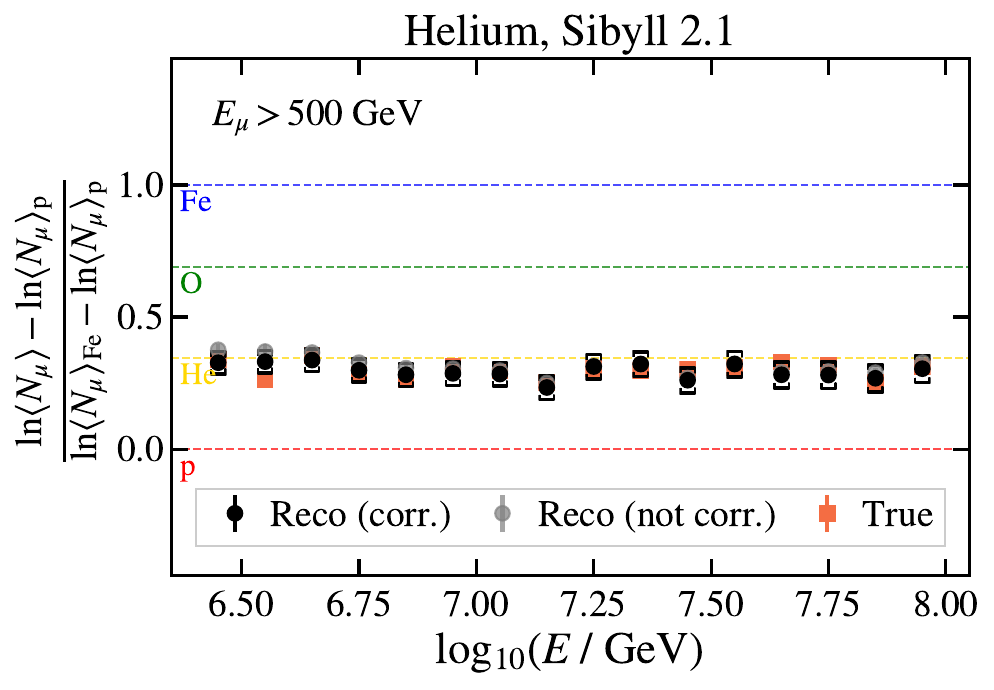}
    
    \includegraphics[width=0.45\textwidth]{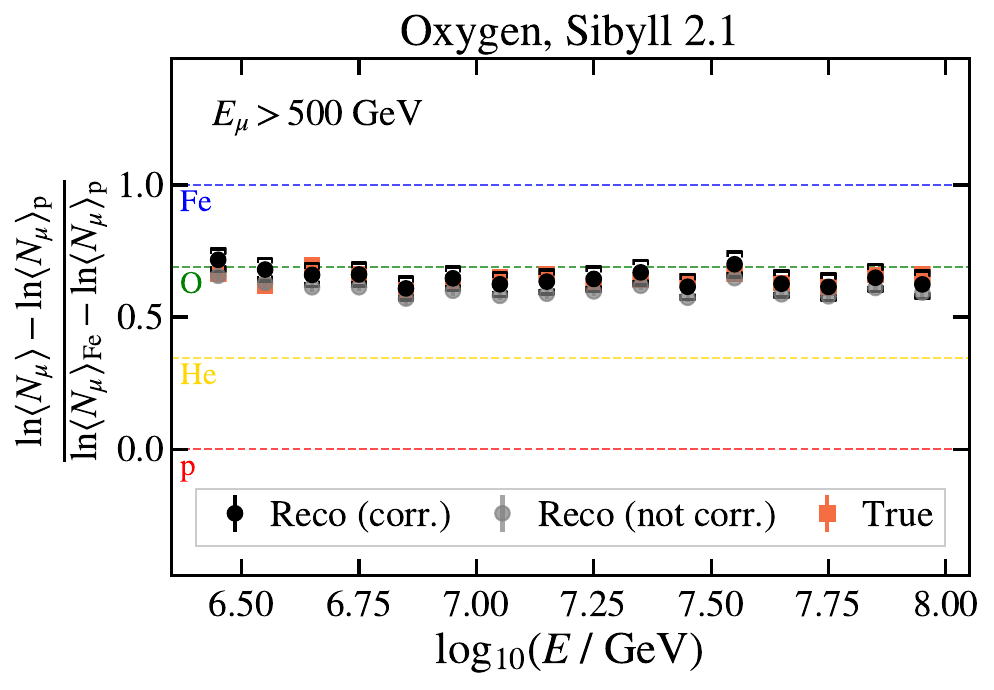}\qquad\includegraphics[width=0.45\textwidth]{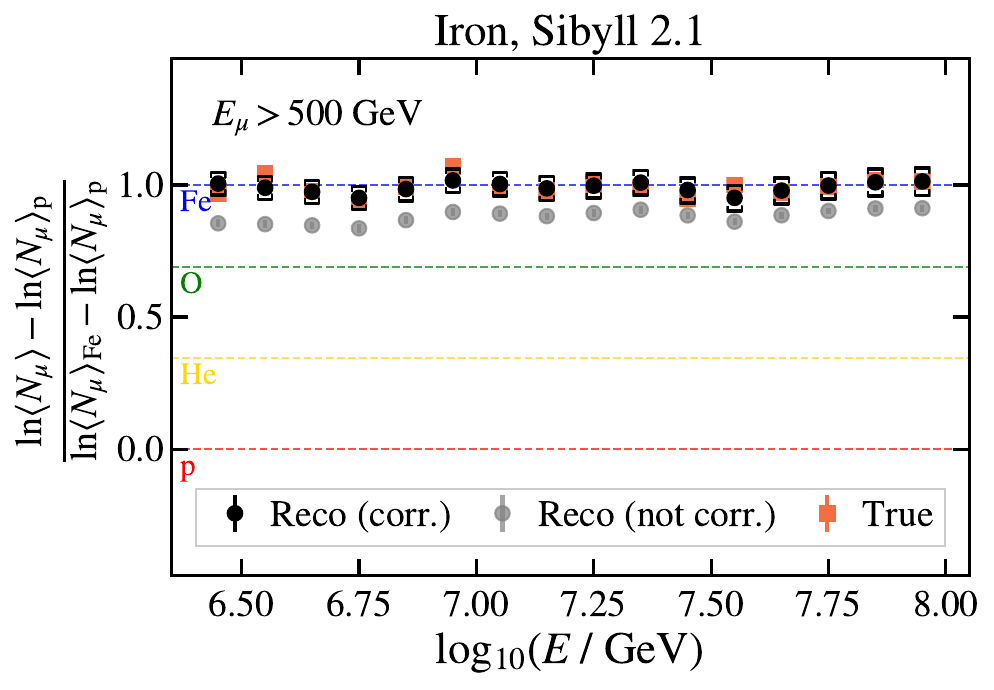}
    \vspace{-0.1cm}
    \caption{Comparison between the initial reconstruction of the average muon number (in bins of reconstructed energy), the corrected results, and the true muon number (in bins of true energy) obtained from air showers simulated with \sibyllpre. The values have been scaled according to \refeq{eq:z}, so that the true value for proton is at zero and the true value for iron is at 1. The brackets show the uncertainty assigned to the final result to account for the typical offsets remaining after the correction. This comparison is shown for four pure composition scenarios.}
    \label{fig:correction_works}
\end{figure*}

The correction factors derived from simulations can be used to unbias the results that will be obtained from experimental data. A complication is that the correction factors are clearly mass dependent, while the cosmic-ray composition at high energies is not precisely known. Rather than assuming a specific composition model to derive corrections, a model-independent approach is applied. It exploits the fact that the muon number itself is a measure of composition, and that the values of the correction factors at a specific energy depend, to a good approximation, linearly on \lnA, with $A$ being the nuclear mass number. This works as follows. The Heitler-Matthews model~\cite{Matthews:2005sd} predicts that the number of muons in an air shower depends on $A$ and the cosmic-ray energy $E$ as
\begin{equation}\label{eq:Nmu_heitler}
    N_\mu (E, A) = A^{1-\beta} \left(\frac{E}{\xi}\right)^\beta,
\end{equation}
with an exponent $\beta < 1$ and $\xi$ a constant.\footnote{In \refapp{app:HE_mu}, we discuss the validity of this relation for the high-energy (\Emugtr) muons in the shower.} This relation inspires the definition of the $z$ value, a common way of representing muon data by scaling it according to expectations from simulations~\cite{EAS-MSU:2019kmv},
\begin{equation}\label{eq:z}
    z = \frac{\ln \langle N_\mu \rangle - \ln \langle N_\mu\rangle_\mathrm{p}}{\ln \langle N_\mu\rangle_\mathrm{Fe} - \ln \langle N_\mu\rangle_\mathrm{p}},
\end{equation}
where $\langle N_\mu\rangle_\mathrm{p}$ and $\langle N_\mu\rangle_\mathrm{Fe}$ are the predictions from proton and iron simulations, respectively. The relation of \refeq{eq:Nmu_heitler} can thus be used to estimate \lnA based on the \Nav derived from the data as $z \approx \lnA / \ln (56)$. With the estimate of \lnA obtained from the initial \Nav estimate, a correction factor can be obtained by linearly interpolating the correction factors derived for proton and iron, $\mathcal{C}_\mathrm{p}$ and $\mathcal{C}_\mathrm{Fe}$, in \lnA, 
\begin{equation}\label{eq:interp}
    \mathcal{C}_{\lnA}(E) = \mathcal{C}_\mathrm{p}(E) + \frac{\mathcal{C}_\mathrm{Fe}(E)-\mathcal{C}_\mathrm{p}(E)}{\ln (56)}\lnA.
\end{equation}
The interpolated correction factor, $\mathcal{C}_{\ln A}$, can then be applied to the initial estimate of \Nav to obtain a corrected estimate. This process is iterated, using the newly corrected \Nav values to estimate the mass composition and in turn derive a new correction factor. After several iterations, the process converges to what is then the final result for the average muon number. The approximate dependence of the correction factors on \lnA has been tested by deriving correction factors for He and O by interpolation (i.e.~ \refeq{eq:interp} for $A=4$ and $16$) and comparing them to the ones derived directly from simulation (which are shown in \reffig{fig:reco_v_true}).

The method has been verified using simulations. An example is shown in \reffig{fig:correction_works}, where results were derived from simulations of a single primary mass which have undergone the whole analysis process. The corrected results can be seen to closely align with the Monte Carlo truth. This was repeated for a variety of possible mass compositions, and the method was found to reproduce the true \Nav well, independently of the injected composition. The typical difference remaining between the corrected and true values after the correction are taken into account in the final result: the muon number is corrected for an average offset of 0.2\%, and a spread of 4\% is included as a systematic uncertainty, shown as the brackets in \reffig{fig:correction_works}.

\begin{figure}
    \centering
    \includegraphics[width=0.9\linewidth, trim = 0 0.8em 0 0, clip]{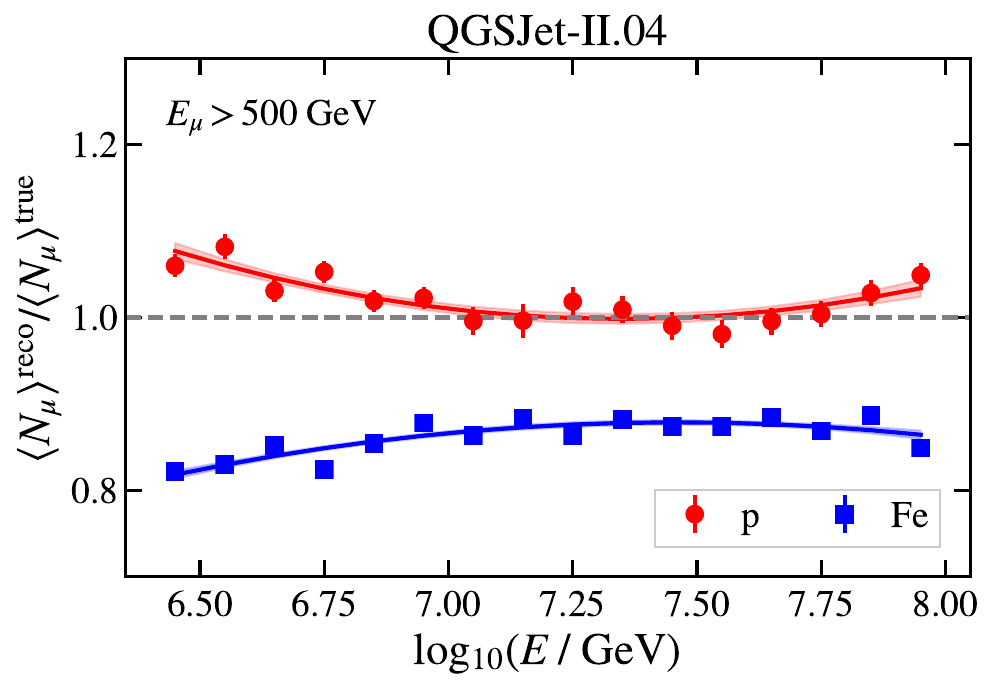}

    \includegraphics[width=0.9\linewidth, trim = 0 0.8em 0 0, clip]{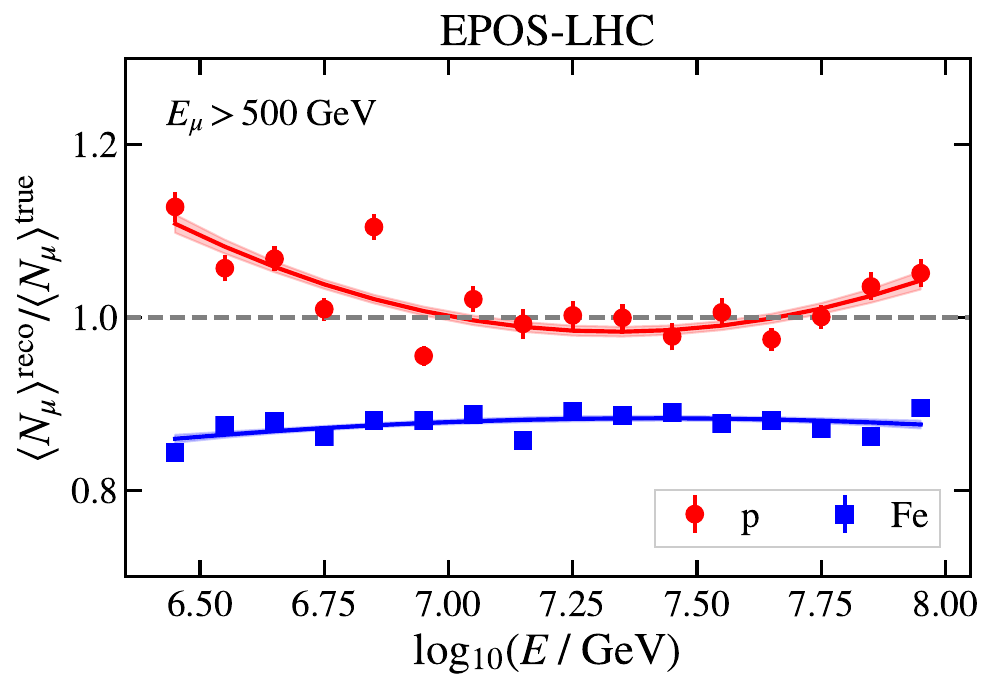}
    \vspace{-0.1cm}
    \caption{Correction factors derived from simulations based on \qgsjet and \epos, equivalent to the \sibyllpre correction factors shown in \reffig{fig:reco_v_true}.}
    \label{fig:correction_factors}
\end{figure}

The correction factors depend on the hadronic interaction model used in the simulations they are derived from. Correction factors are obtained for \qgsjet and \epos simulations in addition to those for \sibyllpre; they are shown in \reffig{fig:correction_factors}. The correction factors effectively remove the dependence of the result on the hadronic model that the initial neural network reconstructions were trained on, and replaces it by a dependence on the model that the correction factors are derived from. They therefore allow us to interpret the experimental data under the assumption of different hadronic interaction models. This is shown in more detail in \refapp{app:checks}, where results are obtained using a neural network trained on \epos, leading to results consistent with the nominal results presented in the following section.

\section{Results and Discussion}\label{sec:results}

\begin{figure}
    \centering
    \includegraphics[width=0.9\linewidth, trim = 0 1em 0 0, clip]{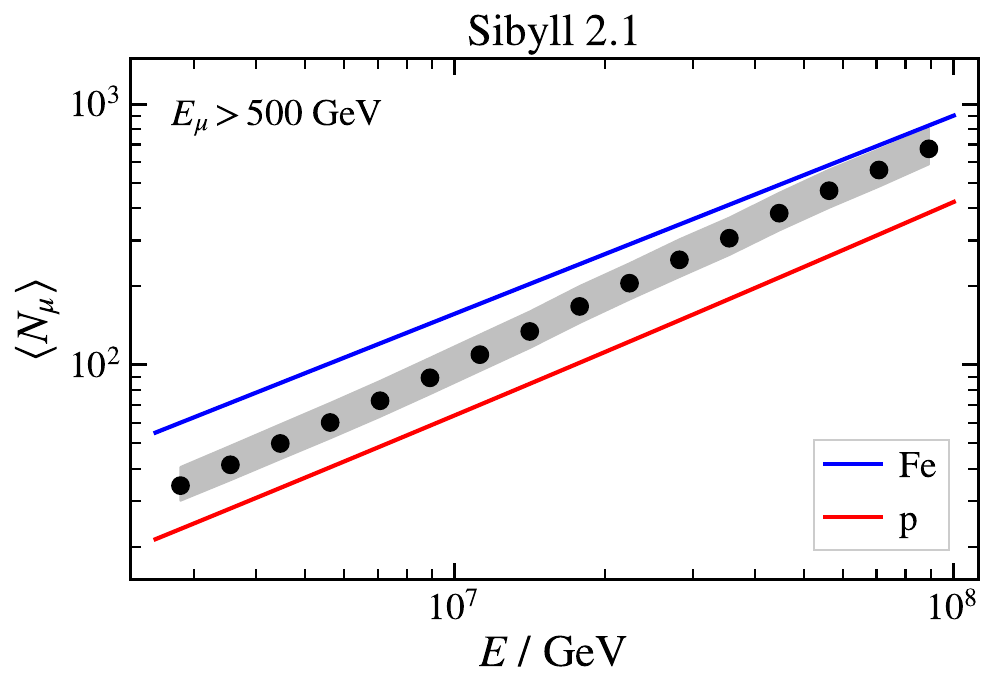}

    \includegraphics[width=0.9\linewidth, trim = 0 1em 0 0, clip]{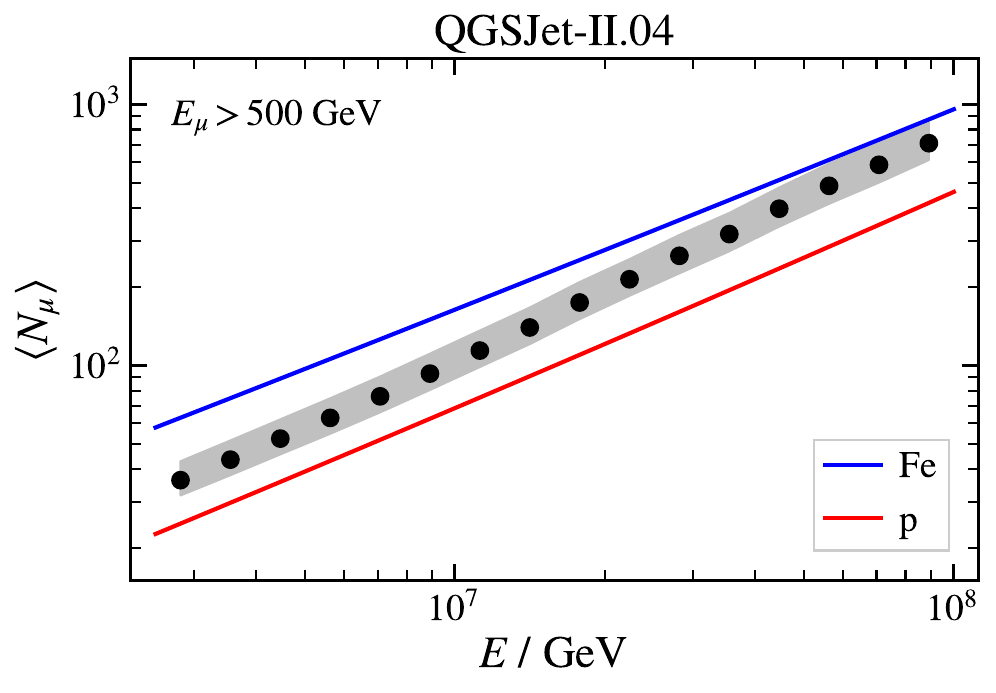}

    \includegraphics[width=0.9\linewidth, trim = 0 1em 0 0, clip]{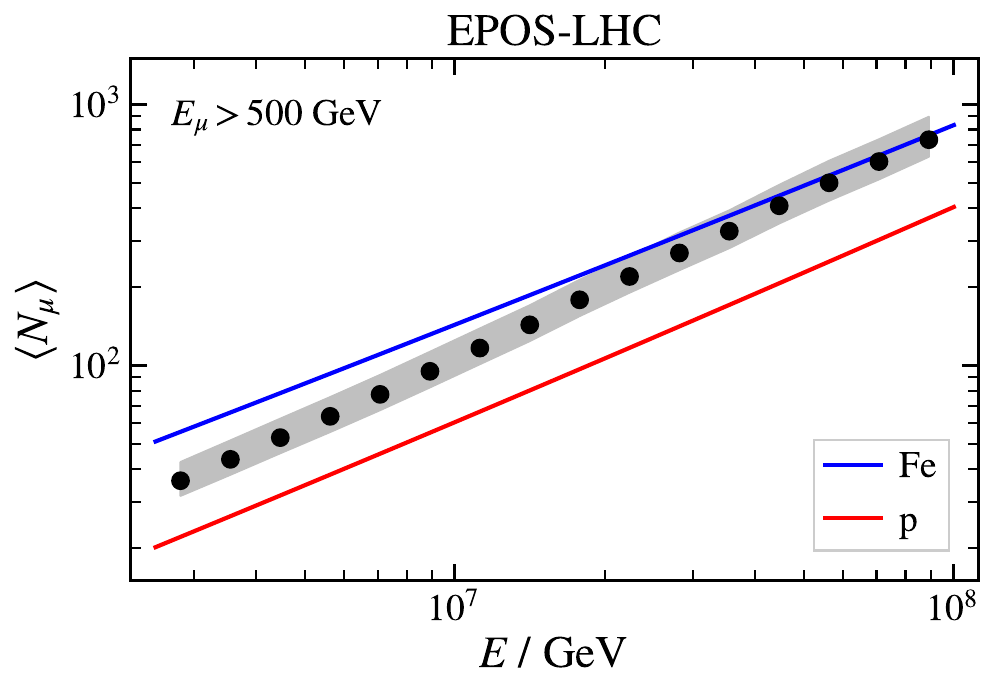}
    \vspace{-0.1cm}
    \caption{Average number of muons with energy greater than \SI{500}{\giga\eV} in near-vertical air showers as a function of the primary cosmic-ray energy obtained using the hadronic interaction models \sibyllpre, \qgsjet, and \epos. The shaded region indicates the systematic uncertainty, statistical uncertainties are not visible. The muon number expected from proton and iron simulations performed with the corresponding hadronic models are shown for comparison. These data are made available in a public data release~\cite{datarelease}.}
    \label{fig:results_N}
\end{figure}

Figure~\ref{fig:results_N} shows the mean number of muons \Nav with energies greater than \SI{500}{\giga\eV} in air showers with energies between \SI{2.5}{\peta\eV} and \SI{100}{\peta\eV}. These results were obtained in a model-dependent way by using the correction factors derived from simulations based on \sibyllpre, \qgsjet, and \epos (shown in \reffigs{fig:reco_v_true}{fig:correction_factors}). Also shown are the predictions of the muon number in proton and iron showers derived from \corsika simulations using the respective hadronic interaction models, as described in \refapp{app:HE_mu}. The shaded area around the points represents the total systematic uncertainties, while statistical uncertainties are too small to be visible in the figures. The central values are bracketed by the proton and iron predictions for all models, and qualitatively indicate that the mass composition becomes heavier with increasing primary energy. 

\begin{figure}
    \centering
    \includegraphics[width=0.9\linewidth, trim = 0 1em 0 0, clip, trim = 0 1em 0 0, clip]{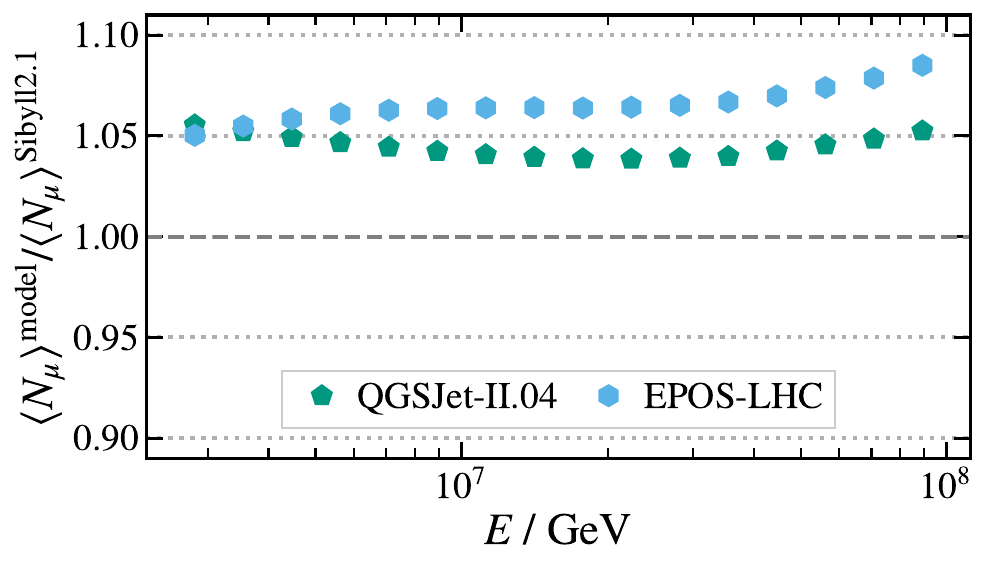}
    \vspace{-0.1cm}
    \caption{Ratio of the muon multiplicity (\Emugtr) results obtained with \qgsjet and \epos to those obtained with \sibyllpre, as shown in \reffig{fig:results_N}.}
    \label{fig:Nmu_ratios}
\end{figure}

\begin{figure}
    \centering
    \includegraphics[width=0.9\linewidth, trim = 0 1em 0 0, clip]{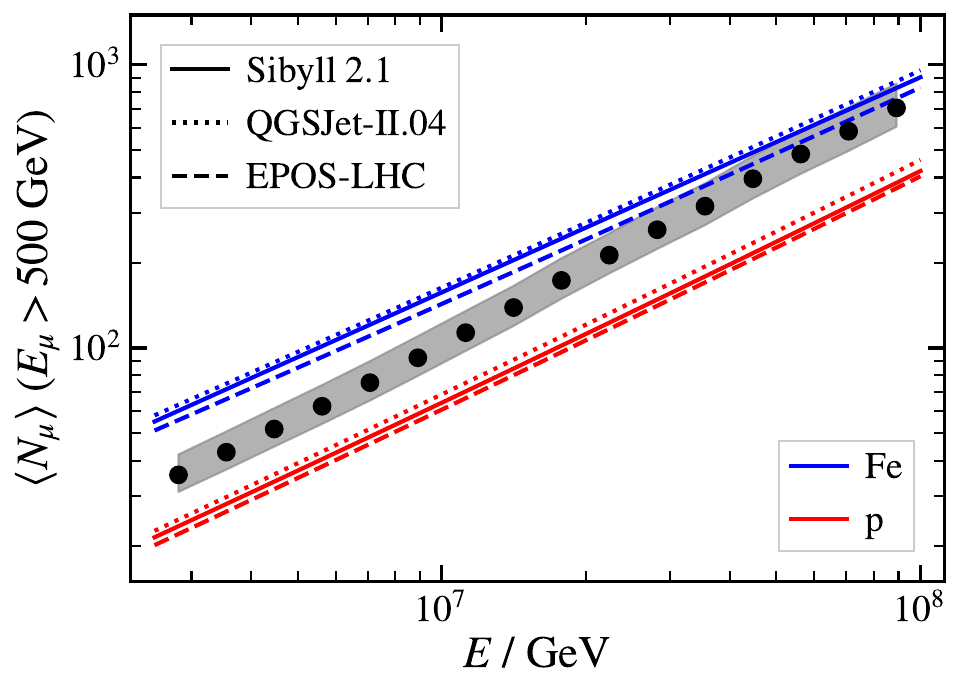}
    \vspace{-0.1cm}
    \caption{Average number of muons with energy greater than \SI{500}{\giga\eV} in near-vertical air showers as a function of primary cosmic-ray energy, averaged over the individual results obtained with \sibyllpre, \qgsjet, and \epos. The systematic uncertainty band includes the deviation of the individual results from the average added in quadrature with the other systematic uncertainties (see text for details). The corresponding simulated muon numbers for proton and iron are shown for comparison. These data are made available in a public data release~\cite{datarelease}.}
    \label{fig:result_average}
\end{figure}

As shown in \reffig{fig:Nmu_ratios}, the muon numbers obtained using \qgsjet and \epos are approximately 5\% higher than the result obtained with \sibyllpre. As this spread is significantly smaller than the total systematic uncertainty (see below), we derive in addition a model-averaged muon multiplicity, shown in \reffig{fig:result_average}. This result includes an additional contribution to the total systematic uncertainty, representing the spread of the model-specific \Nav results around the average. The average result is plotted together with the proton and iron predictions based on the different hadronic models. The \Nav predictions for \qgsjet and \epos are, respectively, about 5\% higher and lower than for \sibyllpre; a more detailed comparison between the predictions is shown in \reffig{fig:MC_pred}. The numerical values for the individual and model-averaged \Nav results are made publicly available in electronic format in \refref{datarelease}.


\begin{figure}
    \centering
    \includegraphics[width=0.9\linewidth, trim = 0 1em 0 0, clip]{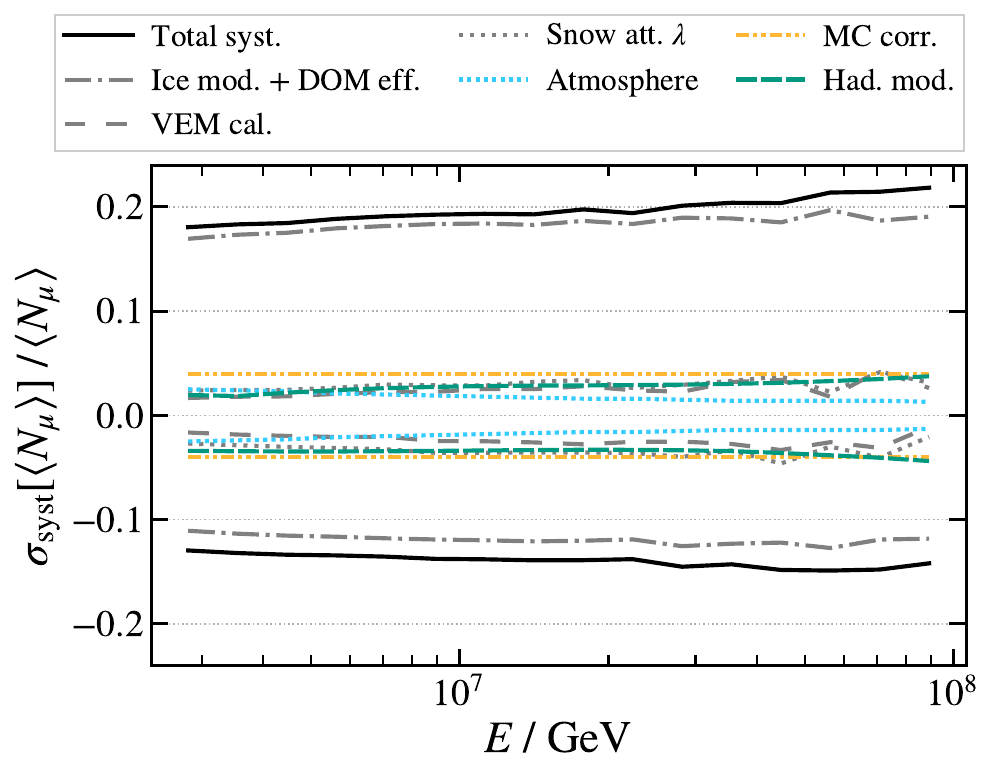}
    \vspace{-0.1cm}
    \caption{Relative size of the systematic uncertainties of the high-energy muon multiplicity measurement. For the model-average result of \reffig{fig:result_average}, all shown uncertainties are included and added in quadrature to obtain the total systematic uncertainty. For the individual model results shown in \reffig{fig:results_N}, all but the hadronic model uncertainty ('Had. mod.') are included.}
    \label{fig:systematics}
\end{figure}

The total systematic uncertainty on the muon number measurement originates from different sources, shown in \reffig{fig:systematics}. Detector uncertainties originate from both IceTop and the in-ice array, and are essentially the same as those described in previous works~\cite{IceCube:2019hmk}. The largest uncertainty is related to the modeling of the properties of the deep ice used in simulations (scattering and absorption of both the bulk ice and the refrozen ice in the drill holes around the DOMs). In \refref{IceCube:2019hmk}, the uncertainties related to the ice model were combined with a 3\% uncertainty on the DOM efficiency to determine the total in-ice light-yield uncertainty for down-going muon bundles. We opt here for a more conservative DOM efficiency uncertainty of 10\%, leading to a total light-yield uncertainty of (+13.4\%, $-$15.7\%). The impact on the result has been determined by repeating the energy-loss reconstruction with a modified light-collection efficiency parameter. For IceTop, an uncertainty of $\pm \SI{0.2}{\m}$ is assumed for the effective attenuation length ($\lambda = \SI{2.25}{\m}$) of air shower particles in  the snow~\cite{IceCube:2012nn, IceCube:2013ftu}. This impacts both the shower size reconstruction, as well as the core and direction reconstruction, which in turn impact the track used in the muon-bundle energy-loss reconstruction. The impact of this uncertainty is evaluated by redoing the whole analysis chain after running an air-shower reconstruction with a different value of $\lambda$. Furthermore, a $\pm 3\%$ uncertainty on the calibration of the VEM charge unit for IceTop was included, based on simulations of the calibration process with different atmospheres, hadronic interaction models, and other systematic variations~\cite{VanOverloop:2011tkr}. The impact of energy-bin migration on the result is included in the definition of the correction factors discussed in \refsec{sec:correction}. An uncertainty of 4\% is assigned to cover any offsets that may remain after the correction procedure. Finally, the difference between the simulated atmosphere (\refsec{sec:datasets}) and the actual atmosphere measured by the AIRS satellite~\cite{AIRS2013}, averaged over the data-taking period, is evaluated. Using the muon-production parametrization from \refref{Gaisser:2021cqh} and the uncertainty on the atmospheric temperature measurements, we conservatively estimate that \Nav might deviate by up to 2.5\% between the simulated and real atmosphere. All uncertainties are added in quadrature and shown as bands in the main results of \reffig{fig:results_N}. For the model-average result shown in \reffig{fig:result_average}, the maximal deviation of the individual results was included as an additional systematic uncertainty, shown as the green lines in \reffig{fig:systematics}. Note that this hadronic interaction model uncertainty only reflects the spread between the models included in this analysis and that other models may produce larger differences.

\begin{figure}
    \centering
    \includegraphics[width=0.9\linewidth, trim = 0 1em 0 0, clip]{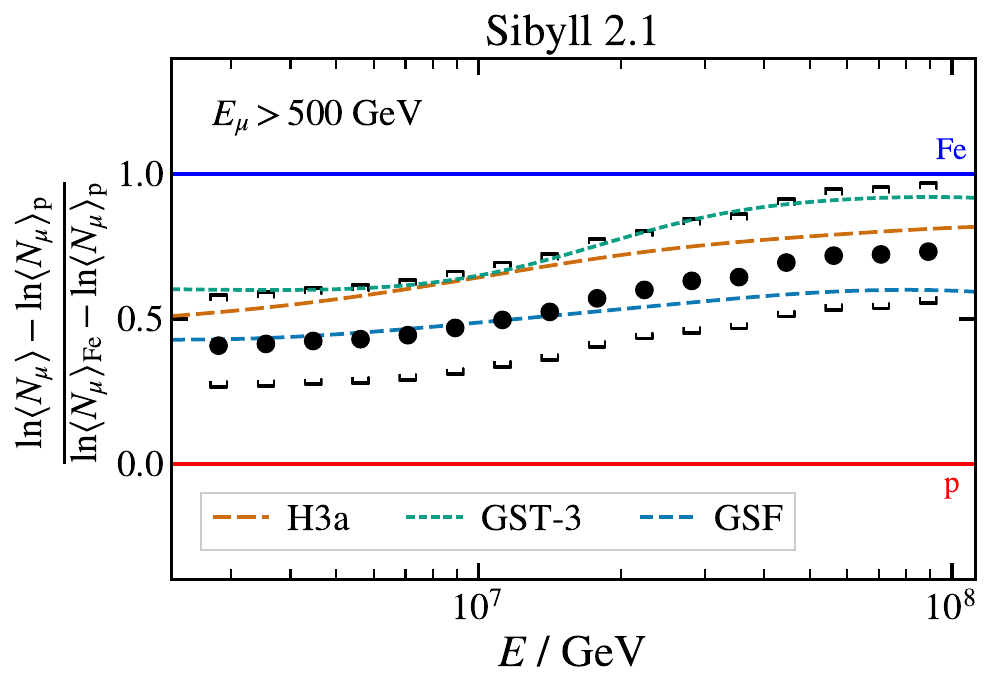}

    \includegraphics[width=0.9\linewidth, trim = 0 1em 0 0, clip]{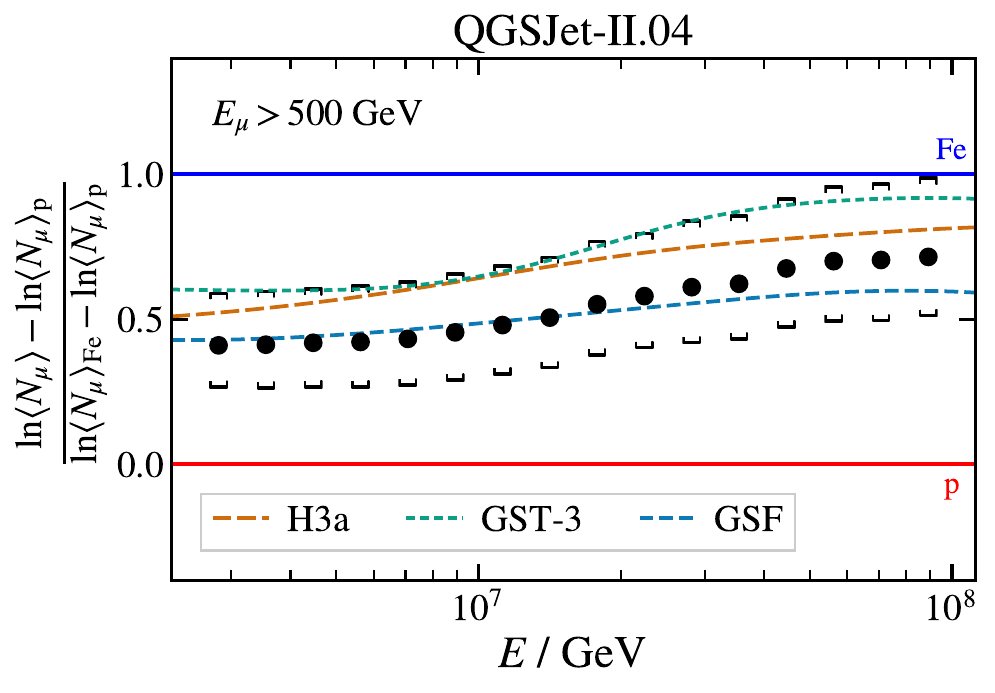}

    \includegraphics[width=0.9\linewidth, trim = 0 1em 0 0, clip]{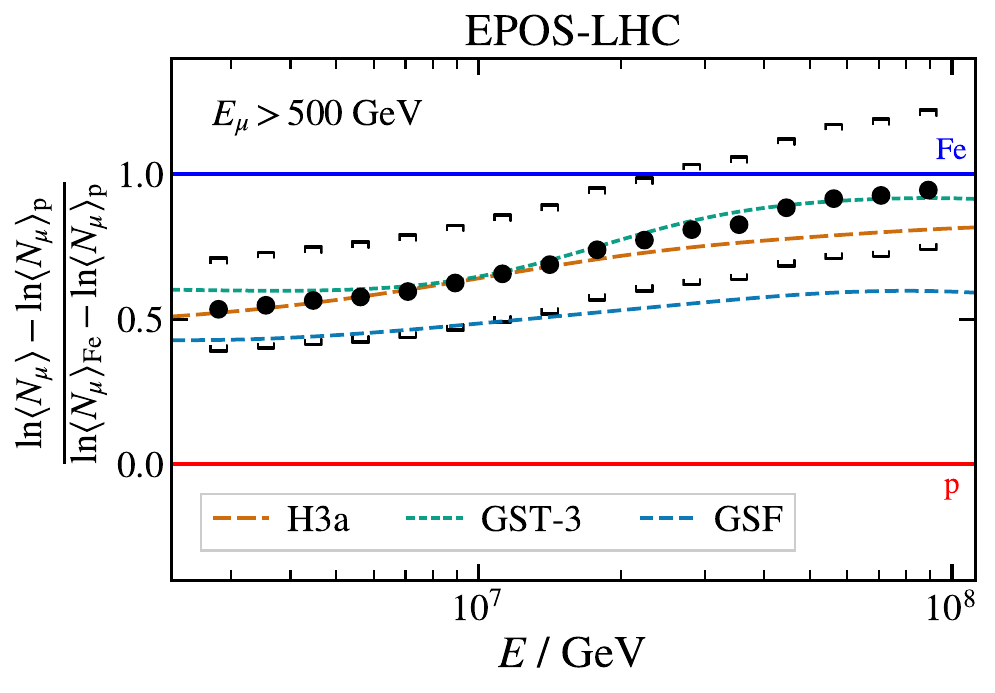}
    \vspace{-0.1cm}
    \caption{Measured muon multiplicities of \reffig{fig:results_N} scaled with respect to the expectations from proton and iron simulations ($z$ values) using different hadronic interaction models, as in \refeq{eq:z}. Systematic uncertainties are indicated by the brackets. The dashed lines display the expected muon multiplicities according to the cosmic-ray flux models H3a, GST, and GSF.}
    \label{fig:results_z}
\end{figure}

Figure~\ref{fig:results_z} shows the $z$-value representation of the results as defined in \refeq{eq:z}, i.e.~the logarithms of the measured high-energy muon number derived using different hadronic interaction models scaled against the expectations from proton and iron simulations using the corresponding model. The result based on \epos indicates a slightly heavier mass composition than the results based on \sibyllpre and \qgsjet, which are very similar in the $z$-values. Note that, in this representation, the difference between the values obtained for the various hadronic interaction models results from the model dependence of both the determination of the \Nav results and of the predictions, which are of similar size (see \reffigs{fig:Nmu_ratios}{fig:MC_pred}). The experimental results are compared to predictions based on three composition models, commonly known as H3a~\cite{Gaisser:2011klf}, GST~\cite{Gaisser:2013bla}, and GSF~\cite{Dembinski:2017zsh}.\footnote{The predictions are calculated as $\Sigma_i f_i \langle N_\mu \rangle_i$, where the sum runs over the different mass groups $i$ included in the model and $f_i$ is the corresponding fraction of the total flux. The expected muon multiplicities $\langle N_\mu \rangle_i$ are the values derived from simulations for $i=(\mathrm{p},\mathrm{Fe})$, and are linearly interpolated in \lnA for intermediate mass groups.} The observed muon number and its evolution with energy generally agrees with the expectations from these models within uncertainties.\footnote{Note that while H3a is independent of IceCube results, the GST and GSF models include IceCube composition results derived from high-energy muons in their fit.}

\begin{figure}
    \centering
    \includegraphics[width=0.9\linewidth, trim = 0 1em 0 0, clip]{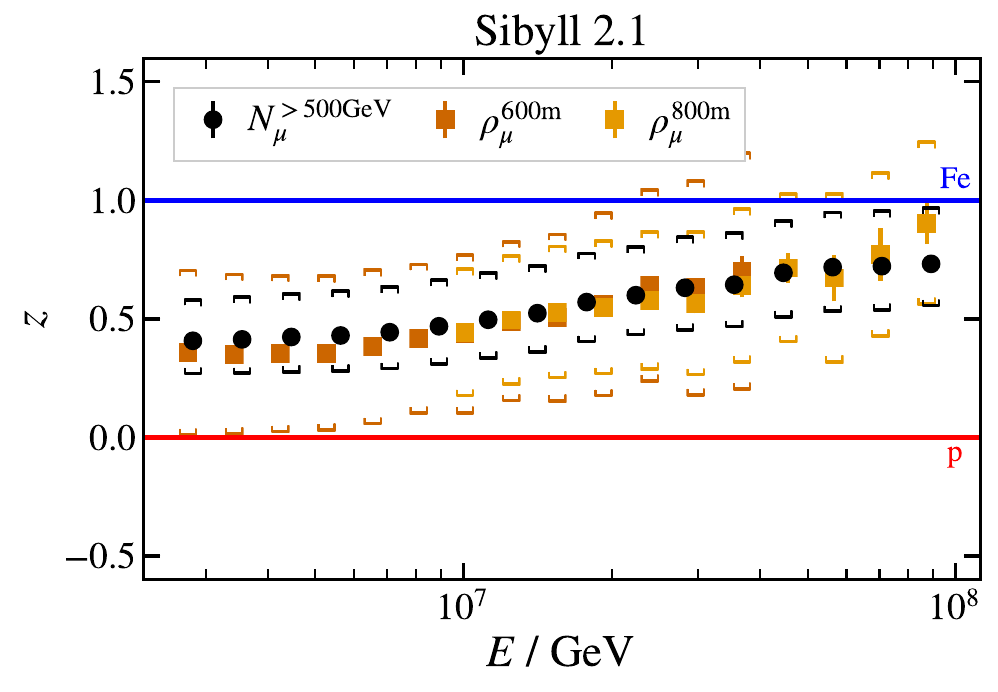}

    \includegraphics[width=0.9\linewidth, trim = 0 1em 0 0, clip]{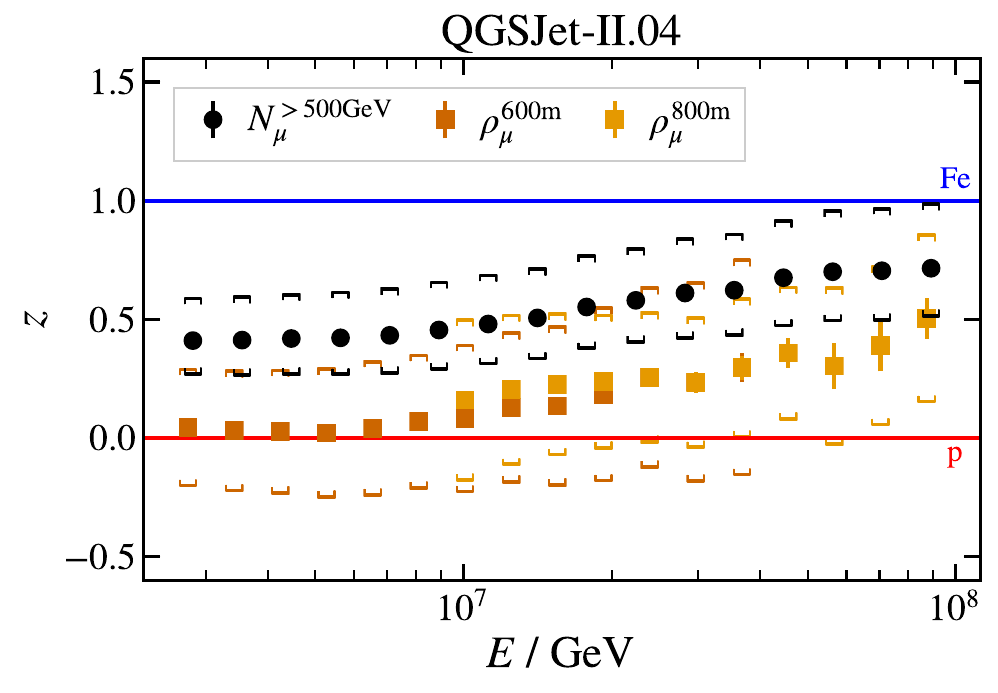}

    \includegraphics[width=0.9\linewidth, trim = 0 1em 0 0, clip]{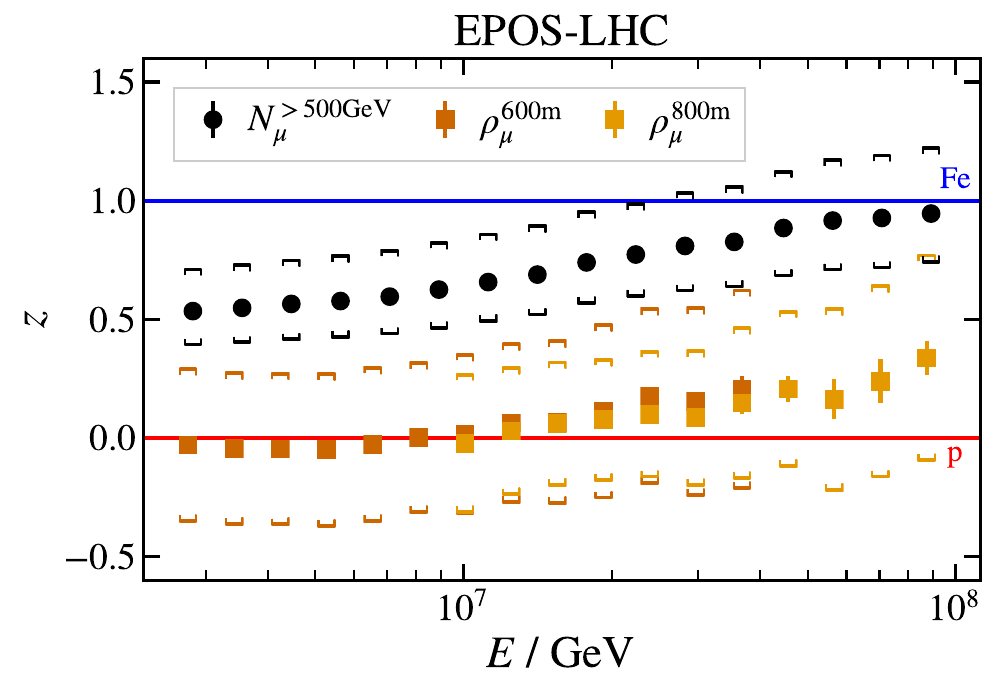}
    \vspace{-0.1cm}
    \caption{Comparison between the TeV muon data obtained in this work ($N_\mu$, black), and the density of GeV muons measured with IceTop at different lateral distances ($\rho_\mu$, shades of orange), represented as $z$-values (\refeq{eq:z}). See the text for details.}
    \label{fig:results_rho}
\end{figure}

It is of interest to compare these results to a muon analysis performed with IceTop alone, mentioned earlier in \refsec{sec:intro}, and described in detail in \refref{IceCubeCollaboration:2022tla}. In that analysis, the density of muons at the surface $\rho_\mu$ was determined at lateral distances of \SI{600}{\m} and \SI{800}{\m} from the shower axis. These are mainly low-energy muons, with a threshold of several \SI{100}{\mega\eV}, commonly referred to as GeV muons. The GeV muon density analysis covers the primary energy range used in the TeV muon analysis presented in this work, and uses the same zenith range ($\cos \theta > 0.95$) and hadronic interaction models. The results of the GeV muon density analysis and the TeV muon multiplicity analysis are shown together in terms of $z$ values in \reffig{fig:results_rho}. If the simulations consistently describe the experimental data, the two results should be consistent, as they are measures of the same primary cosmic-ray flux arriving at Earth. This is the case for the \sibyllpre results, where we observe excellent agreement over the entire energy range. The increased production of low-energy muons in the post-LHC models \qgsjet and \epos results in the GeV muon measurement being closer to expectations for lighter primaries.
The tension with the TeV muon result is strongest for \epos, where the $z$-values obtained from the GeV and TeV muon measurements are outside each other's uncertainty bands over nearly the entire energy range. We note that there are also preliminary indications for inconsistencies related to the slope of the lateral charge distribution observed in IceTop, most prominently for \sibyllpre simulations, as reported in \refref{IceCube:2021ixw}.

Several tests have been performed to ensure the robustness of the TeV muon result. First, it was checked that the nominal result is consistent with results obtained when the energy and muon multiplicity reconstructions are performed separately based on only IceTop and in-ice information respectively. Second, it was confirmed that a neural network trained on simulations based on a different hadronic model produces consistent results after deriving and applying the correction factors. Third, the analysis was performed replacing the neural-network estimated primary energy by a simple energy estimate based on the shower size \Sone, derived for an analysis of the energy spectrum, reported in \refref{IceCube:2013ftu}. This relation was also used in the GeV muon density analysis. The results obtained with this approach are consistent with the nominal result of \reffig{fig:results_N}. More details about these consistency tests are given in \refapp{app:checks}.

\section{Conclusion}\label{sec:conclusion}

We have presented a measurement of the average number of muons with energies greater than \SI{500}{\giga\eV} in near-vertical air showers observed in both IceTop and the IceCube in-ice array for cosmic-ray energies between \SI{2.5}{\peta\eV} and \SI{100}{\peta\eV}. The results were derived using the hadronic interaction models \sibyllpre, \qgsjet, and \epos. They were compared with predictions from simulations, based on the respective hadronic models, and were found to agree within uncertainties with expectations from realistic primary cosmic-ray flux models.

The high-energy muon measurement was in addition compared to an earlier measurement of the density of low-energy muons at the surface performed with IceTop alone, by scaling both to expectations of proton and iron simulations. Represented in this fashion, both results should be in agreement and should point toward the same underlying mass composition.
This is the case for the \sibyllpre results. 
Interpreting the low-energy muon density data with the post-LHC models \qgsjet and \epos yields, however, lighter mass compositions than the TeV muon measurement. 
While both measurements are still consistent within systematic uncertainties for \qgsjet, this is not the case for \epos. This tension indicates that the data are not described consistently by the simulations using this model, and confirms again the challenges in correctly describing the development of extensive air showers, as discussed in \refsec{sec:intro}. 
In addition, we note that preliminary work also indicates a possible inconsistency between the muon measurements and the slope of the lateral charge distribution in IceTop, in particular for \sibyllpre~\cite{IceCube:2021ixw}.

The measurement of high-energy muons has considerable potential for future studies of air-shower development and tests of hadronic interaction models, including newer iterations of the models included in this work. The measurement of the average muon multiplicity can be extended toward higher primary energies, closing the gap to measurements at ultrahigh energy observatories, and a larger zenith range, probing different muon energies and atmospheric depths. It is of interest to study the distribution of high-energy muons in more detail, e.g.~by considering in addition to the average number also its fluctuations~\cite{Cazon:2018gww, PierreAuger:2021qsd}.  More precise combined measurements of the low- and high-energy muon content of air showers will be important toward resolving the Muon Puzzle, as they probe the muon energy spectrum, which differs between hadronic interaction models~\cite{Riehn:2019jet}. Of particular interest are measurements of the correlation between the low- and high-energy muons on an event-by-event basis; while this work already presented an event-by-event reconstruction of the high-energy muon number, a technique to get an improved low-energy muon estimator based on separately fitting the electromagnetic and muonic lateral distribution functions at the surface is under development~\cite{IceCube:2023suf}.

The plans for IceCube-Gen2 and its surface array will increase the opening angle for possible coincident measurements between the surface and deep detector~\cite{IceCube-Gen2:2020qha}. The planned surface radio antennas would offer sensitivity to the depth of shower maximum $X_\mathrm{max}$ in addition to the muon measurements, enabling more stringent tests of the consistency of hadronic models, an important condition for the unambiguous interpretation of indirect cosmic-ray measurements~\cite{Coleman:2022abf}.

\begin{acknowledgements}
The IceCube collaboration acknowledges the significant contributions to this manuscript from Stef Verpoest.
The authors gratefully acknowledge the support from the following agencies and institutions:
USA {\textendash} U.S. National Science Foundation-Office of Polar Programs,
U.S. National Science Foundation-Physics Division,
U.S. National Science Foundation-EPSCoR,
U.S. National Science Foundation-Office of Advanced Cyberinfrastructure,
Wisconsin Alumni Research Foundation,
Center for High Throughput Computing (CHTC) at the University of Wisconsin{\textendash}Madison,
Open Science Grid (OSG),
Partnership to Advance Throughput Computing (PATh),
Advanced Cyberinfrastructure Coordination Ecosystem: Services {\&} Support (ACCESS),
Frontera and Ranch computing project at the Texas Advanced Computing Center,
U.S. Department of Energy-National Energy Research Scientific Computing Center,
Particle astrophysics research computing center at the University of Maryland,
Institute for Cyber-Enabled Research at Michigan State University,
Astroparticle physics computational facility at Marquette University,
NVIDIA Corporation,
and Google Cloud Platform;
Belgium {\textendash} Funds for Scientific Research (FRS-FNRS and FWO),
FWO Odysseus and Big Science programmes,
and Belgian Federal Science Policy Office (Belspo);
Germany {\textendash} Bundesministerium f{\"u}r Bildung und Forschung (BMBF),
Deutsche Forschungsgemeinschaft (DFG),
Helmholtz Alliance for Astroparticle Physics (HAP),
Initiative and Networking Fund of the Helmholtz Association,
Deutsches Elektronen Synchrotron (DESY),
and High Performance Computing cluster of the RWTH Aachen;
Sweden {\textendash} Swedish Research Council,
Swedish Polar Research Secretariat,
Swedish National Infrastructure for Computing (SNIC),
and Knut and Alice Wallenberg Foundation;
European Union {\textendash} EGI Advanced Computing for research;
Australia {\textendash} Australian Research Council;
Canada {\textendash} Natural Sciences and Engineering Research Council of Canada,
Calcul Qu{\'e}bec, Compute Ontario, Canada Foundation for Innovation, WestGrid, and Digital Research Alliance of Canada;
Denmark {\textendash} Villum Fonden, Carlsberg Foundation, and European Commission;
New Zealand {\textendash} Marsden Fund;
Japan {\textendash} Japan Society for Promotion of Science (JSPS)
and Institute for Global Prominent Research (IGPR) of Chiba University;
Korea {\textendash} National Research Foundation of Korea (NRF), Chung-Ang University Research Grant;
Switzerland {\textendash} Swiss National Science Foundation (SNSF).
\end{acknowledgements}

\bibliographystyle{apsrev4-2}
\bibliography{main}

\clearpage

\appendix

\section{Monte Carlo predictions of TeV muon multiplicity}\label{app:HE_mu}

\begin{figure}
    \centering    
    \includegraphics[width=0.9\linewidth, trim = 0 1em 0 0, clip]{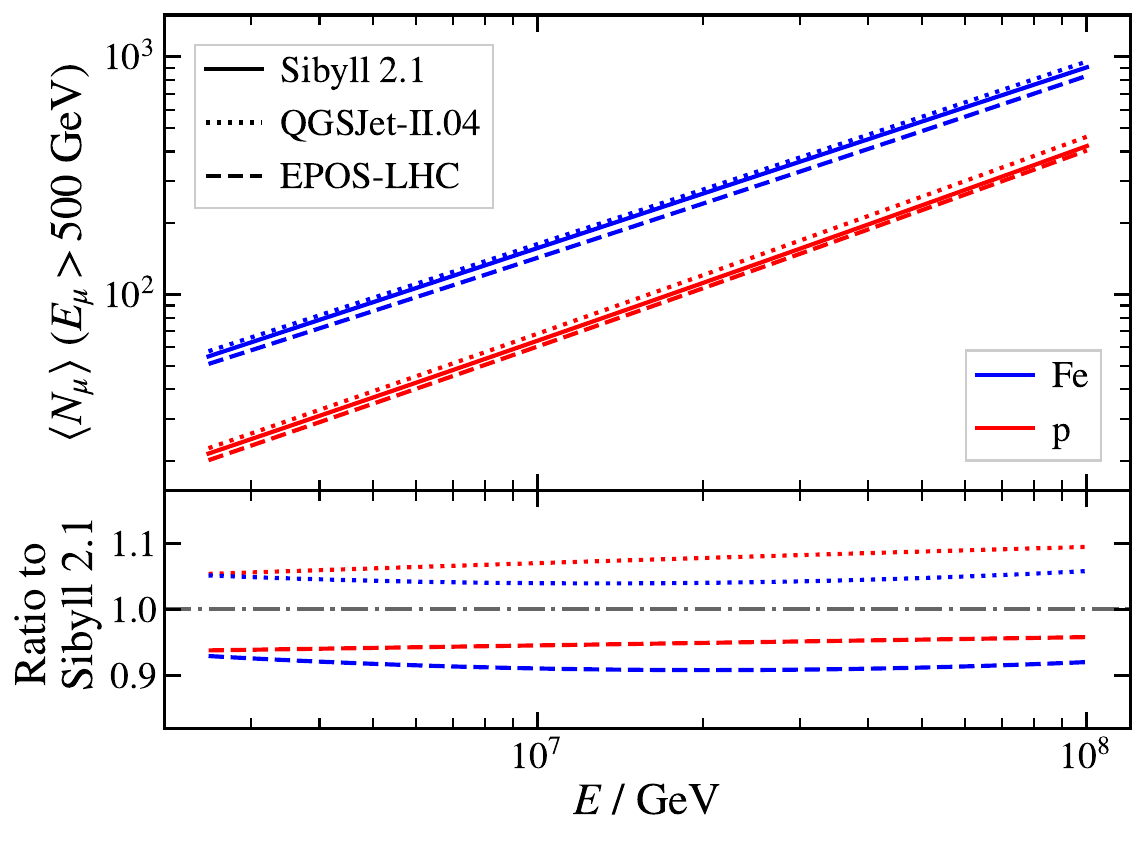}
    \caption{Average multiplicity of muons with an energy above \SI{500}{\giga\eV} in near-vertical air showers ($\cos \theta > 0.95$) as a function of cosmic-ray energy. The lines are quadratic fits to predictions obtained from \corsika simulations for proton and iron primaries based on different hadronic interaction models.}
    \label{fig:MC_pred}
\end{figure}

To obtain predictions of the number of high-energy muons with negligible statistical uncertainty from simulations for comparison with the final analysis results, dedicated high-statistics \corsika-only simulations were produced, tracking particles down to \SI{500}{\giga\eV} only. Figure~\ref{fig:MC_pred} shows the predictions for \Nav (\Emugtr) at an observation level of \SI{2837}{\m} obtained from simulations of near-vertical proton and iron showers using different hadronic interaction models. \qgsjet predicts the highest number of muons, \epos the lowest. 

The Heitler-Matthews model of air-shower development, combined with the superposition assumption for showers initiated by nuclei of mass number $A$, predicts that the total number of muons in the shower depends on the cosmic-ray energy $E$ and mass $A$ as
\begin{equation}\label{eq:Nmu_heitler_app}
    N_\mu (E, A) = A^{1-\beta} \left(\frac{E}{\xi}\right)^\beta
\end{equation}
where $\beta \approx 0.9$ is related to the multiplicity of the hadronic interactions and $\xi$ is a constant called the critical energy~\cite{Matthews:2005sd}. From \reffig{fig:MC_pred}, it can be seen that the high-energy muon number has a similar dependence on energy. Fitting the slope, the value of $\beta$ is found to be about 0.82 for proton and 0.76 for iron. According to \refeq{eq:Nmu_heitler_app}, the exponent also determines the difference between \Nmu for proton and heavier nuclei. From the difference between the proton and iron predictions at fixed $E$, $\beta$ is derived to be about 0.77 at the lower primary energies increasing to about 0.82 at the higher end.\footnote{This smaller value of $\beta$ compared to the $\sim 0.9$ corresponds to the proton-iron difference for high-energy muons being larger than for the total number of muons at the surface. The high-energy muons are thus more sensitive to the mass composition.} The fact that the $\beta$ values derived from the energy dependence and the mass dependence are numerically close implies that the superposition model also holds well for the description of the high-energy muons. It is interesting to note that this different behavior compared to the total muon number makes the interpretation of a high-energy muon measurement less sensitive to possible offsets in the energy scale of an experiment~\cite{EAS-MSU:2019kmv}. Whereas a 10\% shift in energy scale would result in a difference of about 0.2 in $z$ (\refeq{eq:z}) for a surface muon measurement, this would cause a shift of only about 0.08 in $z$ for the high-energy muons.

While the relation between \Nav and \EE is close to a straight line on the logarithmic plot, \reffig{fig:MC_pred}, the residuals after a linear fit show small deviations with a quadratic behavior on the level of 2\%. For the accuracy of the calculations of e.g.~the final $z$-values of \reffig{fig:results_z}, a quadratic term was added to the fit describing $\langle N_\mu \rangle(E)$. 

\section{Neural network consistency checks}\label{app:checks}

In this section of the Appendix, various checks are presented showing the robustness of the analysis. 

\subsection{Separate \boldmath{$E$} and \boldmath{$N_\mu$} networks}

\begin{figure*}
    \centering
    \includegraphics[width=0.36\textwidth]{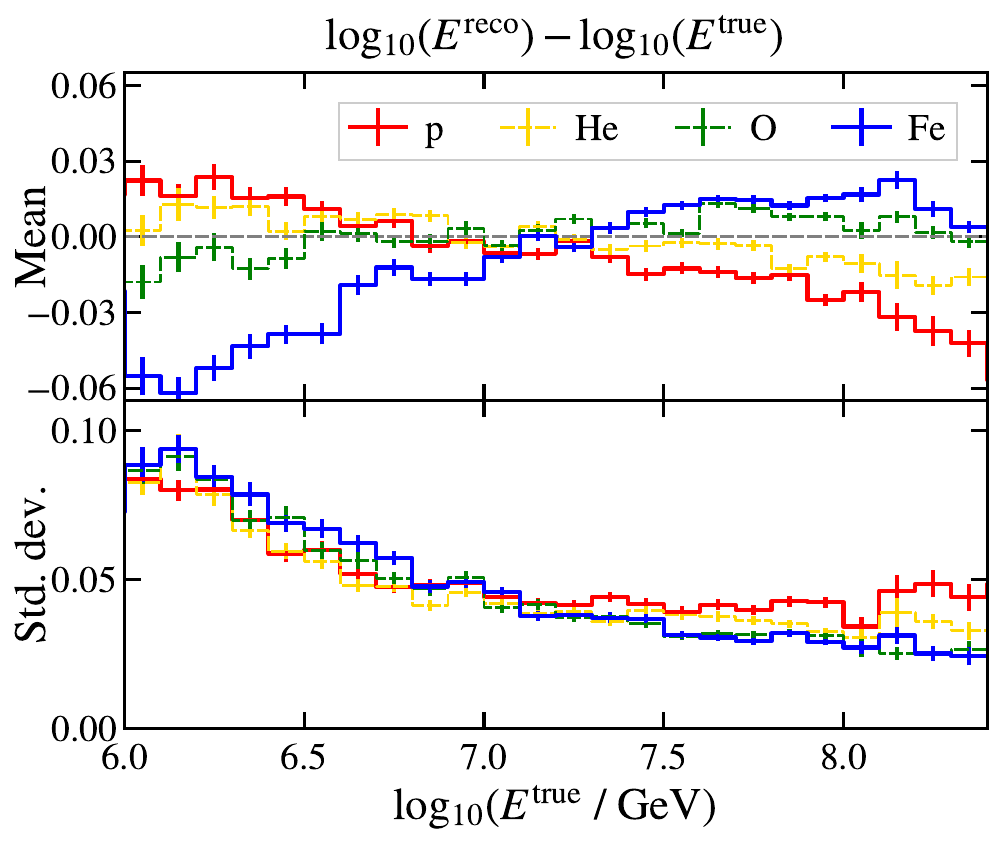}\qquad\qquad\qquad\includegraphics[width=0.36\textwidth]{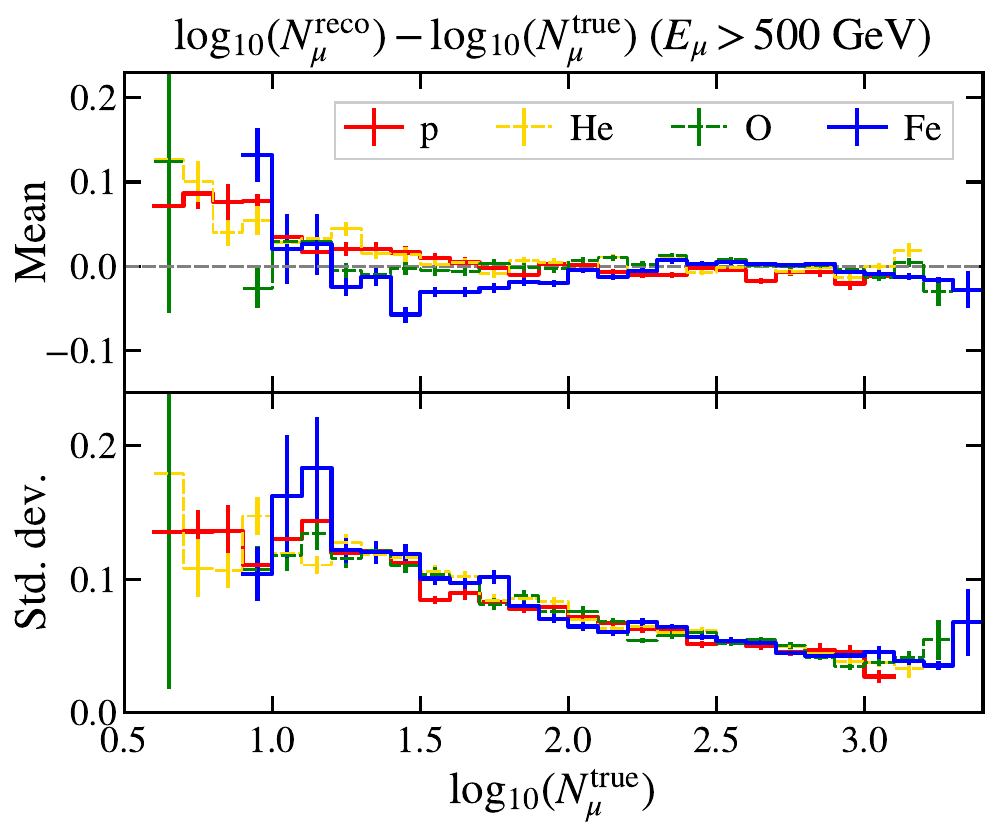}
    \caption{Bias and resolution of the energy reconstruction based only on IceTop information (left) and the TeV muon number reconstruction based only on in-ice information (right).}.
    \label{fig:separate_NN_reso}
\end{figure*}

The neural network model in this analysis uses three inputs to reconstruct the primary cosmic-ray energy \EE and the high-energy muon number \Nmu (\Emugtr): the shower size \Sone and zenith angle \tta from IceTop, and a vector of energy losses from the IceCube in-ice array. To ensure that consistent results are obtained when the primary energy and the muon number reconstructions are performed completely independently, two different neural networks were trained. One uses \Sone and \tta to reconstruct \EE, the other uses the energy losses to reconstruct \Nmu. The reconstruction quality for these networks is shown in \reffig{fig:separate_NN_reso}. Compared to the performance of the combined neural network, shown in \reffig{fig:NN_reso}, the \EE reconstruction has an increased mass-dependent bias, indicating that the combined neural network uses the composition information present in the muon-bundle signal to improve the \EE reconstruction. The \Nmu neural network, which does not use any surface information, shows, on the other hand, a reduced mass dependence in the bias.

    

\begin{figure}
    \centering
    \includegraphics[width=0.8\linewidth, trim = 0 0.8em 0 0, clip]{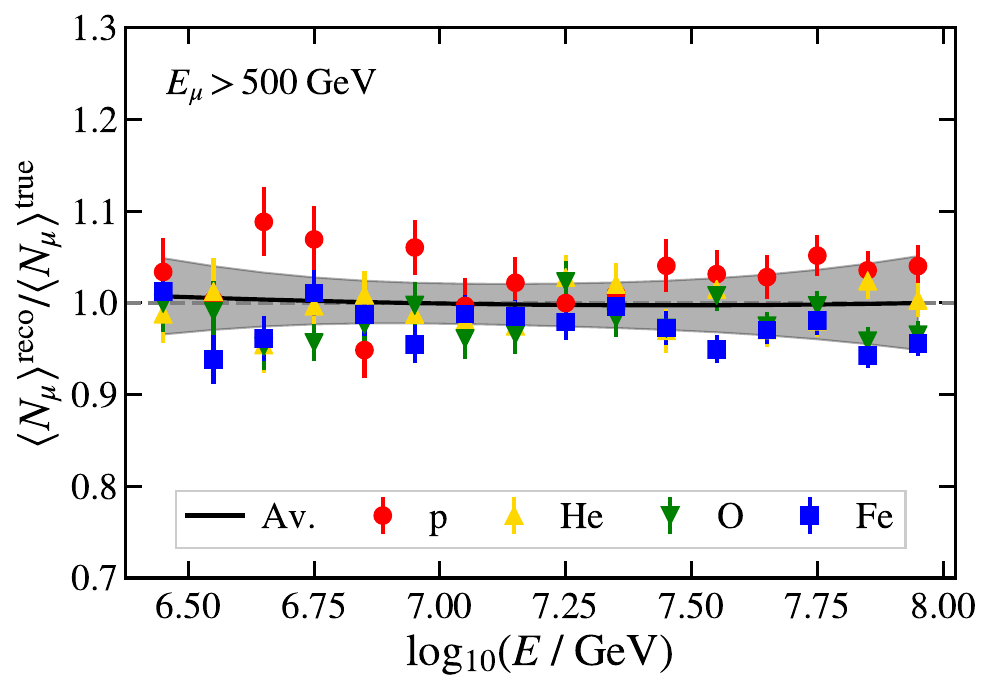}
    \caption{Ratio of reconstructed and true average muon number obtained from the reconstructions with separate neural networks for \EE and \Nmu. The black line defines the average of fits to proton and iron, the shaded area shows the uncertainty assigned to cover the unknown mass composition.}
    \label{fig:separate_NN_corr}
\end{figure}

As before, correction factors can be determined for each primary type by comparing the reconstructed and true average muon number \Nav. The resulting biases are shown in \reffig{fig:separate_NN_corr}. They are found to be smaller than those in \reffig{fig:reco_v_true}, however, they do not have a monotone relation with the mass of the primary. Therefore, an iterative correction where one interpolates between the proton and iron correction factors as in \refsec{sec:correction} cannot be applied; this is the main reason for using the combined neural network in the main body of this work. To derive results based on the separate reconstruction, a single correction factor is derived as the average between the proton and iron correction factors, and the spread between the two is added as an extra systematic uncertainty on the result obtained with this correction factor.

\begin{figure}
    \centering
    \includegraphics[width=0.85\linewidth, trim = 0 1em 0 0, clip]{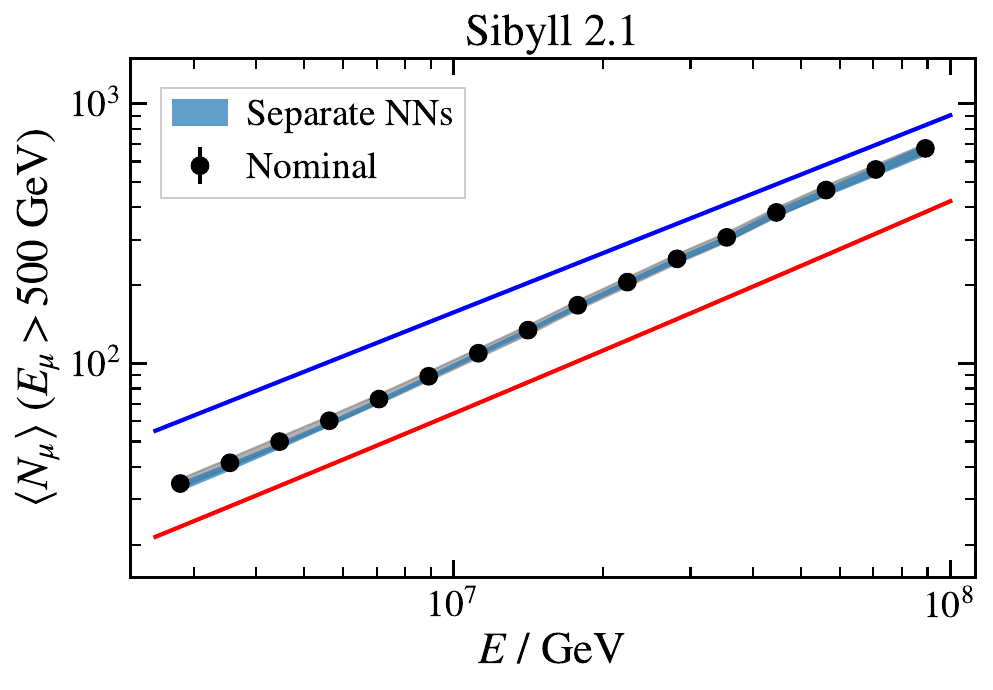}
    \caption{Comparison between the nominal result of the analysis and the result obtained using separate neural networks with only IceTop input for the energy reconstruction and only IceCube in-ice array input for the muon multiplicity reconstruction.}
    \label{fig:separate_NN}
\end{figure}

The result obtained with the separate neural networks and the average p-Fe correction factor is shown in \reffig{fig:separate_NN} compared to the nominal result presented in \reffig{fig:results_N}. The two are in agreement for the entire energy range.

We note that, while no iterative correction as presented in \refsec{sec:correction} can be applied here, the smaller biases observed in \reffig{fig:separate_NN_corr} as compared to \reffig{fig:reco_v_true} would lead to uncertainties of a similar size on the final result.

\subsection{Training on different hadronic models}

\begin{figure}
    \centering
    \includegraphics[width=0.85\linewidth, trim = 0 1em 0 0, clip]{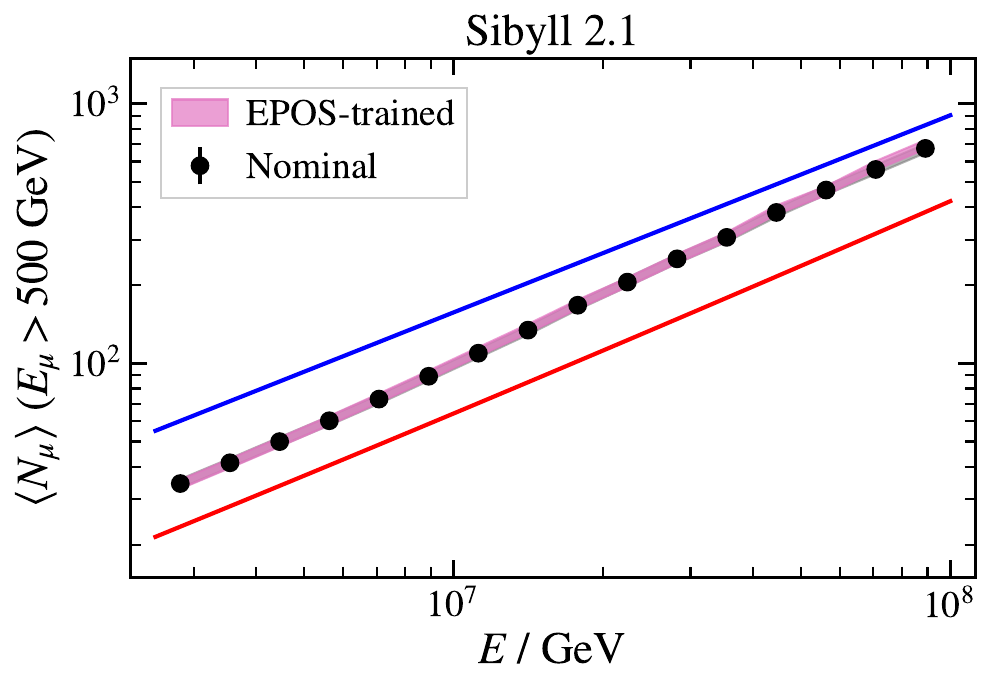}
    \caption{Comparison between the nominal result of the analysis and the result obtained using a neural network that was trained on simulations based on \epos. In both cases, correction factors were derived from \sibyllpre simulations.}
    \label{fig:epos_trained}
\end{figure}

The nominal results presented in the main body of this paper are obtained using a neural network trained on \sibyllpre after which correction factors derived from \sibyllpre, \qgsjet, and \epos are applied. We show here that the results depend only on the model that the correction factors are derived from and are robust against changes in the model that the neural network is trained on. To this end, a neural network was trained on \epos simulations. A correction factor was then derived by applying this network to \sibyllpre simulations. The result obtained from experimental data with this combination of neural network and correction factors is shown in \reffig{fig:epos_trained}. The result is consistent with the nominal result obtained from training a network on \sibyllpre and then deriving a correction factor from \sibyllpre. Similarly, training the network on \epos and deriving correction factors based on \epos gives consistent results as using the \sibyllpre-trained network and the \epos-derived correction factors from \refsec{sec:analysis}.

\subsection{Simple shower-size energy estimator}

\begin{figure}
    \centering
    \includegraphics[width=0.85\linewidth, trim = 0 1em 0 0, clip, trim = 0 1em 0 0, clip]{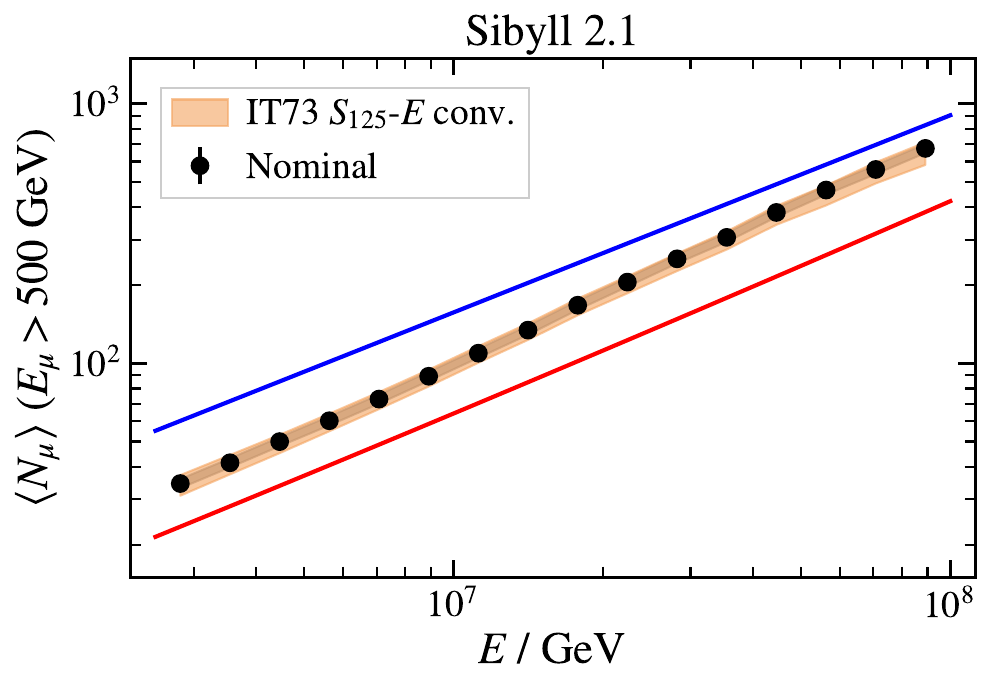}
    \caption{Comparison between the nominal result of the analysis and the result obtained when the energy is reconstructed with a simple conversion from \Sone derived for the IceTop-73 energy-spectrum analysis (see text for details).}
    \label{fig:Ereco_check}
\end{figure}

Previous IceTop analyses typically estimate the primary cosmic-ray energy through a conversion function based on the shower size \Sone defined for a specific range of zenith angles. The IceTop analysis of the surface muon density, of which some results were shown in \refsec{sec:results}, uses for its energy estimate a conversion function derived for the energy-spectrum analysis performed with IceTop-73, before the completion of the full IceTop detector, as described in \refref{IceCube:2013ftu}. The energy estimate in the high-energy muon analysis of this work comes from a neural network which also gets input from the in-ice detector.  Earlier in this Appendix it was already confirmed that obtaining an energy estimate from a separate neural network based only on \Sone and \tta from IceTop does not produce a significantly different result. Here, we test the effect of using the simple \Sone conversion function used in the GeV muon density analysis, while leaving the \Nmu reconstruction unchanged. In \reffig{fig:Ereco_check}, it is observed that the resulting muon measurement is again consistent with the nominal result.

\clearpage

\end{document}